\documentclass[twocolumn,floatfix,aps,prb,shownopacs,footinbib,superscriptaddress,longbibliography]{revtex4-2}
\usepackage{amsmath}
\usepackage{amsfonts}
\usepackage{amssymb}
\usepackage{graphicx}
\usepackage{epsfig}
\usepackage{color}
\usepackage{empheq}
\usepackage{braket}
\usepackage{epstopdf}
\setcitestyle{super}

\begin{document}
\title{Observation of the scaling dimension of fractional quantum Hall anyons}

\author{A.~Veillon}
\affiliation{Universit\'e Paris-Saclay, CNRS, Centre de Nanosciences et de Nanotechnologies, 91120, Palaiseau, France}
\author{C.~Piquard}
\affiliation{Universit\'e Paris-Saclay, CNRS, Centre de Nanosciences et de Nanotechnologies, 91120, Palaiseau, France}
\author{P.~Glidic}
\affiliation{Universit\'e Paris-Saclay, CNRS, Centre de Nanosciences et de Nanotechnologies, 91120, Palaiseau, France}
\author{Y.~Sato}
\affiliation{Universit\'e Paris-Saclay, CNRS, Centre de Nanosciences et de Nanotechnologies, 91120, Palaiseau, France}
\author{A.~Aassime}
\affiliation{Universit\'e Paris-Saclay, CNRS, Centre de Nanosciences et de Nanotechnologies, 91120, Palaiseau, France}
\author{A.~Cavanna}
\affiliation{Universit\'e Paris-Saclay, CNRS, Centre de Nanosciences et de Nanotechnologies, 91120, Palaiseau, France}
\author{Y.~Jin}
\affiliation{Universit\'e Paris-Saclay, CNRS, Centre de Nanosciences et de Nanotechnologies, 91120, Palaiseau, France}
\author{U.~Gennser}
\affiliation{Universit\'e Paris-Saclay, CNRS, Centre de Nanosciences et de Nanotechnologies, 91120, Palaiseau, France}
\author{A.~Anthore}
\affiliation{Universit\'e Paris-Saclay, CNRS, Centre de Nanosciences et de Nanotechnologies, 91120, Palaiseau, France}
\affiliation{Université Paris Cité, CNRS, Centre de Nanosciences et de Nanotechnologies, F-91120, Palaiseau, France}
\author{F.~Pierre}
\affiliation{Universit\'e Paris-Saclay, CNRS, Centre de Nanosciences et de Nanotechnologies, 91120, Palaiseau, France}

\begin{abstract}
Unconventional quasiparticles emerging in the fractional quantum Hall regime\cite{Wen_BookQFT_2004,Jain_BookCompositeFermions_2007} present the challenge of observing their exotic properties unambiguously. 
Although the fractional charge of quasiparticles has been demonstrated since nearly three decades\cite{Goldman_FracQAntiDot_1995,de-Picciotto_es3_1997,Saminadayar_FracE1s3_1997}, the first convincing evidence of their anyonic quantum statistics has only recently been obtained\cite{Nakamura_Anyon1s3_2020,Bartolomei_Cross_1tiers_2020} and, so far, the so-called scaling dimension that determines the quasiparticles’ propagation dynamics remains elusive.
In particular, while the non-linearity of the tunneling quasiparticle current should reveal their scaling dimension, the measurements fail to match theory, arguably because this observable is not robust to non-universal complications\cite{Rosenow_NonUnivQPCfqhe_2004,Papa_anomTDOSinteraction_2004,Shtanko_SN2s3ExtParam_2014,Dolcetto_TDOSextendedQPC_2012,Snizhko_ScaldimFano_2015}.
Here we expose the scaling dimension from the thermal noise to shot noise crossover, and observe an agreement with expectations.
Measurements are fitted to the predicted finite temperature expression involving both the quasiparticles scaling dimension and their charge\cite{Snizhko_ScaldimFano_2015,Schiller_ScaldimFano_2022}, in contrast to previous charge investigations focusing on the high bias shot noise regime\cite{Heiblum_FracChargeMeas_2010}.
A systematic analysis, repeated on multiple constrictions and experimental conditions, consistently matches the theoretical scaling dimensions for the fractional quasiparticles emerging at filling factors $\nu=1/3$, $2/5$ and $2/3$. 
This establishes a central property of fractional quantum Hall anyons, and demonstrates a powerful and complementary window into exotic quasiparticles.
\end{abstract} 
\maketitle

Exotic quasiparticles could provide a path to protected manipulations of quantum information\cite{Nayak_TopoRMP_2008}.
Yet their basic features are often challenging to ascertain experimentally.
The broad variety of quasiparticles emerging in the regimes of the fractional quantum Hall effect constitutes a prominent illustration. 
These are characterized by three unconventional properties\cite{Wen_BookQFT_2004,Jain_BookCompositeFermions_2007}: \textit{(i)} their charge $e^*$ is a fraction of the elementary electron charge $e$, \textit{(ii)} their anyonic quantum statistics is different from that of bosons and fermions, and \textit{(iii)} the dynamical response to their injection or removal along the propagative edge channels is peculiar, ruled by a `scaling dimension' $\Delta$ different from the trivial $\Delta=1/2$ of non-interacting electrons.
In the simplest Laughlin quantum Hall states, at filling factors $\nu=1/(2n+1)$ ($n\in\mathbb{N}$), the fractional anyon quasiparticles have a charge $e^*=\nu e$, an exchange phase $\theta=\nu\pi$ and a scaling dimension $\Delta=\nu/2$. 
Despite four decades of uninterrupted investigations of the quantum Hall physics, experimental confirmations of the predicted scaling dimension remain lacking, including for Laughlin fractions.

Such a gap may appear surprising since $\Delta$ plays a role in most transport phenomena across quantum point contacts (QPC), the basic building block of quantum Hall circuits.
Indeed, the elementary tunneling process itself consists in the removal of a quasiparticle on one side of a QPC and its reinjection on the other side, whose time correlations are set by $\Delta$\cite{Wen_BookQFT_2004,Jain_BookCompositeFermions_2007}.
In Luttinger liquids, the quasiparticles' scaling dimension is related to the interaction strength, also referred to as the interaction parameter $K$, which notably determines the non-linear $I-V$ characteristics across a local impurity\cite{Giamarchi_TLLbook_2004}.
Consequently, the knowledge of $\Delta$ is often a prerequisite to connect a transport observable with a property of interest. 
Furthermore, as straightforwardly illustrated in the Hong-Ou-Mandel setup\cite{Jonckheere_HOMdelta_2023,Iyer_FiniteWidthAnyons_2024,Thamm_finitewidthanyons_2024}, $\Delta$ naturally rules time controlled manipulations of anyons, which are required in the perspective of topologically protected quantum computation based on braiding\cite{Nayak_TopoRMP_2008}.
In this work, the scaling dimension of fractional quantum Hall quasiparticles is disclosed from the thermal noise to shot noise crossover, as recently proposed\cite{Snizhko_ScaldimFano_2015,Schiller_ScaldimFano_2022}. 
The observed good agreement with universal predictions establishes experimentally the theoretical understanding and completes our picture of the exotic fractional quantum Hall anyons.

\vspace{\baselineskip}
{\noindent\textbf{Characterizing exotic quasiparticles.}}\\
\noindent
The first unconventional property of quantum Hall quasiparticles that has been established is their fractional charge $e^*$.
Consistent values were observed by multiple experimental approaches\cite{Goldman_FracQAntiDot_1995,de-Picciotto_es3_1997,Saminadayar_FracE1s3_1997,Reznikov_es5_1999,Martin_SetFracChargeFQHE_2004,Dolev_nu5s2SNes4_2008,Venkatachalam_SetChargees4Nu5s2_2011,Kapfer_FQHE_2019,Bisognin_PATes3_2019,Roosli_fracCB_2021}, with the main body of investigations based on shot noise measurements across a QPC. 
In this case, the scaling dimension can be canceled out, leaving only $e^*$, by focusing on the ratio between shot noise and tunneling current (the Fano factor) at high bias voltages.
The non-standard braiding statistics of fractional quasiparticles turned out more challenging to observe. 
Convincing evidences were obtained only recently, through Aharonov-Bohm interferometry\cite{Nakamura_Anyon1s3_2020,Nakamura_Anyon2s5_2023} as well as from noise measurements in a `collider' geometry\cite{Bartolomei_Cross_1tiers_2020,Glidic_Collider2023,Ruelle_2s5_2023,Lee_BunchingOrBraiding_2023}.
Note that whereas the latter strategy is particularly versatile, the noise signal is also entangled with the scaling dimension\cite{Rosenow_Collider_2016,Lee_NonAbelianCollider_2022,Iyer_FiniteWidthAnyons_2024,Thamm_finitewidthanyons_2024}, which complicates a quantitative determination of the anyon exchange phase $\theta$\cite{Glidic_Collider2023}.
Finally, the quasiparticles scaling dimension was previously investigated through measurements of the non-linear current-voltage characteristics of a QPC\cite{Roddaro_WBSscaldim_2004,Radu_TDOS5s2_2008,Baer_TDOS5s2_2014}.
However, no reliable value of $\Delta$ could be obtained for the fractional quasiparticles of the quantum Hall regime (see Ref.~\citenum{Anthore_QuSimTLL_2018} for an observation in a circuit quantum simulator, and Ref.~\citenum{Cohen_UnivCLL1s3to1_2023} for a good match on the $I(V)$ of tunneling electrons across a $(\nu=1)-(\nu=1/3)$ interface).
Indeed, the $I-V$ characteristics is generally found at odds with the standard model of a chiral Luttinger liquid with a local impurity (see e.g.\ Refs.~\citenum{Chang_LL_2003,Heiblum_FracChargeMeas_2010,Jain_BookCompositeFermions_2007,Roddaro_WBSscaldim_2004} and references therein).
Most often a fit is impossible, or only by introducing extra offsets and with unrealistic values for $e^*$ and $\Delta$\cite{Radu_TDOS5s2_2008,Lin_TDOS5s2_2012,Baer_TDOS5s2_2014,Ruelle_2s5_2023}. 

\begin{figure}[t!]
	\centering
	\includegraphics[width=\columnwidth]{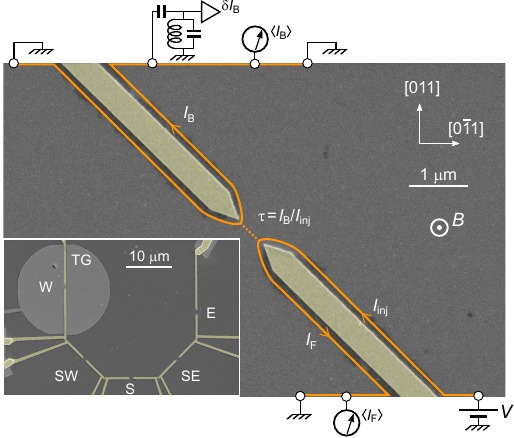}
	\caption{\textbf{Experimental setup.} 
Electron-beam micrographs of the measured Ga(Al)As device.
QPCs are formed in the 2DEG by applying a negative voltage to the metallic gates colored yellow.
The sample includes five QPCs (E, SE, S, SW and W) along different crystallographic orientations (inset, see Extended Data Fig.~\ref{fig:SIsample} for larger-scale images).
Among those, QPC$_\mathrm{W}$ differs by the presence of closely surrounding metallic gate (TG) extending over a 10\,$\mu$m radius (see Extended Data Fig.~\ref{fig:SIQPCW} for close-up images).
Quasiparticle tunneling takes place between chiral quantum Hall edge channels shown as orange lines with arrows (main panel, QPC$_\mathrm{SW}$).
The auto-correlations in back-scattered (tunneling) current $\braket{\delta I_\mathrm{B}^2}$ are measured for all QPCs.
QPC$_\mathrm{E}$ also includes a noise amplification chain for the forward current fluctuations $\delta I_\mathrm{F}$ (not shown), hence allowing for the additional measurements of $\braket{\delta I_\mathrm{F}^2}$ and $\braket{\delta I_\mathrm{B} \delta I_\mathrm{F}}$.
}
	\label{fig:sample}
\end{figure}

The puzzling $I-V$ situation motivated multiple theoretical investigations.
A simple possible explanation for the data-theory mismatch is that the shape of the QPC potential, and therefore the quasiparticle tunneling amplitude, is impacted by external parameters, such as an electrostatic deformation induced by a change in the applied bias voltage, the temperature or the tunneling current itself\cite{Shtanko_SN2s3ExtParam_2014}. 
Other possible non-universal complications include an energy-dependent tunneling amplitude\cite{Dolcetto_TDOSextendedQPC_2012}, additional edge modes either localized\cite{Rosenow_NonUnivQPCfqhe_2004} or propagative\cite{Yang_TDOSwithEdgeReconstruct_2003,Ferraro_AnomTunnelNeutralModes_2008}, and Coulomb interactions between different edges\cite{Papa_anomTDOSinteraction_2004}. 
In this context, the scaling dimension was connected to different, arguably more robust proposed observables such as delta-$T$ noise\cite{Rech_deltaTnoiseFQHE_2020,Zhang_deltaTnoiseFQHE_2022}, thermal to shot noise crossover\cite{Snizhko_ScaldimFano_2015,Schiller_ScaldimFano_2022} and thermal Fano factor\cite{Ebisu_deltaGQvsDelta_2022}. 

A proven strategy to cancel out non-universal behaviors consists in considering a well-chosen ratio of transport properties, as illustrated by the Fano factor $F$ successfully used to extract $e^*$.
Recently, it was proposed that the same $F$ could also give access to the quasiparticles' scaling dimension, when focusing on the lower bias voltage regime where the crossover between thermal noise and shot noise takes place\cite{Snizhko_ScaldimFano_2015,Schiller_ScaldimFano_2022}.
As further detailed later-on, the predicted evolution of $F$ along the crossover exhibits a markedly different width and overall shape depending on the value of $\Delta$.

The present investigation implements the characterization of the scaling dimension from the Fano factor crossover on four different quantum Hall quasiparticles: \textit{(i)} the $e^*=e/3$ quasiparticles, observed at $\nu=1/3$\cite{de-Picciotto_es3_1997} and along the outer edge channel of conductance $e^2/3h$ at $\nu=2/5$\cite{Reznikov_es5_1999,Kapfer_FQHE_2019}, of predicted $\Delta=1/6$\cite{Wen_BookQFT_2004}; \textit{(ii)} the $e^*=e/5$ quasiparticles observed along the inner edge channel of conductance $e^2/15h$  at $\nu=2/5$\cite{Reznikov_es5_1999,Kapfer_FQHE_2019}, of predicted $\Delta=3/10$\cite{Wen_BookQFT_2004,Kane_ImpScattFQHE_1995}; \textit{(iii)} the $e^*=e/3$ quasiparticles observed at $\nu=2/3$\cite{Bid_2s3noisyplateau_2009,Bisognin_PATes3_2019}, of predicted $\Delta=1/3$\cite{Kane_ImpScattFQHE_1995}; \textit{(iv)} the electrons at $\nu=3$ of trivial $\Delta=1/2$.
See Methods for further details on the predictions.

\begin{figure*}[tbh]
	\centering
	\includegraphics[width=180mm]{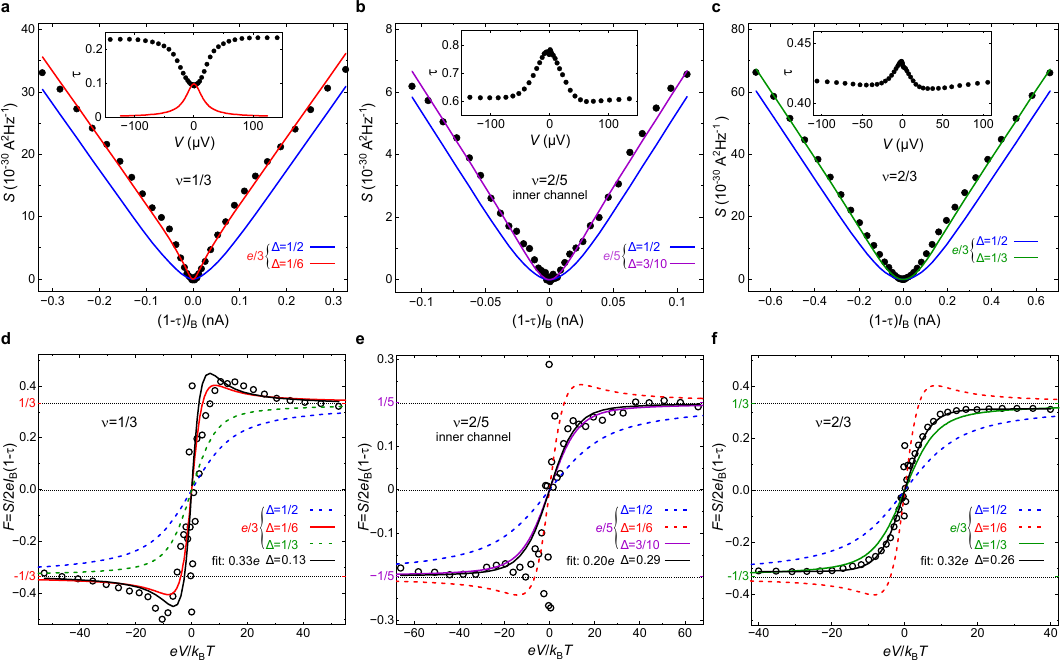}
	\caption{\textbf{Thermal to shot noise crossover.}
Data-theory comparison at $\nu=1/3$ (a,d), on the inner $e^2/15h$ channel of $\nu=2/5$ (b,e) and at $\nu=2/3$ (c,f).
  \textbf{a,b,c,} Excess noise $S$ versus normalized tunnel current $(1-\tau)I_\mathrm{B}$.
  Symbols are measurements (a,d: QPC$_\mathrm{W}$ with $V_\mathrm{tg}=50$\,mV at $T\simeq30.85$\,mK; b,c,e,f: QPC$_\mathrm{E}$ at $T\simeq30.7$\,mK), with a standard error on the noise of $1\times10^{-31}\,\mathrm{A}^2/$Hz, smaller than symbols size.
  Blue lines are phenomenological predictions of Eq.~\ref{eq:Sexc1s2} ($\Delta=1/2$, predicted $e^*$).
  Red, purple and green lines are predictions of Eq.~\ref{eq:SexcDelta} (predicted $\Delta$ and $e^*$).
  Inset: $\tau(V)$ measurements are shown as symbols. 
  The $\nu=1/3$ prediction (red line) differs strongly from these.
  \textbf{d,e,f,} Fano factor $F\equiv S/2eI_\mathrm{B}(1-\tau)$ vs $eV/k_\mathrm{B}T$. 
  Measurements (symbols) agree best with the predictions of Eq.~\ref{eq:SexcDelta} computed using the predicted quasiparticle scaling dimension $\Delta$ (colored continuous lines) than using the electron scaling dimension $1/2$ (blue dashed lines), both assuming the predicted $e^*$.
  Black continuous lines are fits using $e^*$ and $\Delta$ as free parameters.
  }
	\label{fig:TtoSNcrossover}
\end{figure*}

\vspace{\baselineskip}
{\noindent\textbf{Experimental implementation.}}\\
\noindent
The measured sample is shown in Fig.~\ref{fig:sample} together with a schematic representation of the setup.
It is nanofabricated from a Ga(Al)As two-dimensional electron gas (2DEG) and immersed in a strong perpendicular magnetic field corresponding to the quantum Hall effect at filling factors $\nu\in\{1/3,2/5,2/3,3\}$.
Lines with arrows display the chiral propagation of the current along the sample edges.
Quantum point contacts are formed in the 2DEG by field effect, within the opening of metallic split gates (yellow).
We characterize a QPC by the gate-controlled transmission ratio $\tau\equiv I_\mathrm{B}/I_\mathrm{inj}$, with $I_\mathrm{B}$ the back-scattered current and $I_\mathrm{inj}$ the incident current along the edge channel under consideration.
The sample includes five QPCs nominally identical except for their orientation and the presence for QPC$_\mathrm{W}$ of a surrounding gate (labeled TG in the inset of Fig.~\ref{fig:sample}).
This surrounding gate allows us to test the possible influence on the scaling dimension of the local 2DEG density, of an enhanced screening of the long-range Coulomb interactions\cite{Papa_anomTDOSinteraction_2004}, and of an increased sharpness of the electrostatic edge confinement potential\cite{Yang_TDOSwithEdgeReconstruct_2003}. 

The noise is measured using two cryogenic amplifiers (one schematically shown). 
The gain of the noise amplification chains, and the electronic temperature within the device, were obtained from their relation to thermal noise (Methods).
One amplifier (top-left in Fig.~\ref{fig:sample}) measures the back-scattered (tunneling) current noise $\braket{\delta I_\mathrm{B}^2}$ for any addressed QPCs.
A second amplifier (not shown) measures the forward current fluctuations $\delta I_\mathrm{F}$ transmitted specifically across QPC$_\mathrm{E}$.
In practice, we focus on the excess noise with respect to zero bias: $S(V)\equiv\braket{\delta I^2}(V)-\braket{\delta I^2}(0)$.

\begin{figure*}[tbh]
	\centering
	\includegraphics[width=180mm]{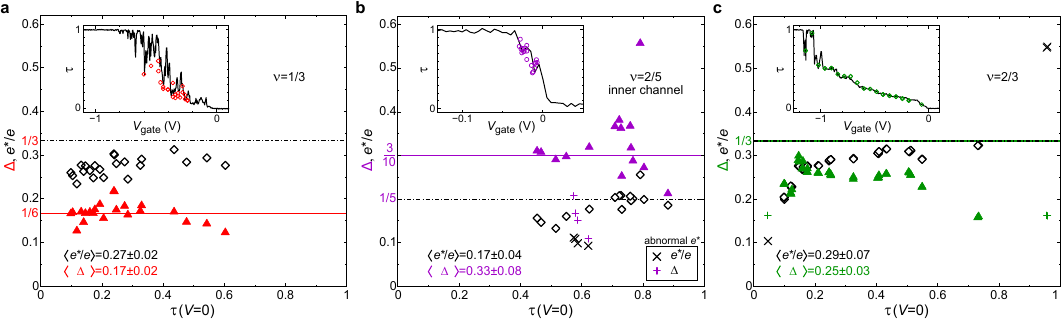}
	\caption{\textbf{Scaling dimension vs QPC tuning.} 
Scaling dimension ($\blacktriangle$) and charge ($\diamond$) obtained along illustrative spans in gate voltage ($V_\mathrm{gate}$) of QPC$_\mathrm{E}$ are plotted vs $\tau(V=0)$ (\textbf{a}, $\nu=1/3$; \textbf{b}, inner channel at $\nu=2/5$; \textbf{c}, $\nu=2/3$).
At $\nu=2/5$ (b) and $\nu=2/3$ (c), a few points are shown as different symbols ($+$,$\times$). They are associated with anomalously low $e^*\lesssim e^*_\mathrm{th}/2$ or high  $e^*\gtrsim 3/2e^*_\mathrm{th}$ charge compared to the predicted charge $e^*_\mathrm{th}$.
Horizontal lines represent the theoretical predictions for $\Delta$ (continuous) and $e^*/e$ (dash-dot).
Inset: Separately measured $\tau(V_\mathrm{gate})$ sweeps (continuous lines), and individual noise measurement tunings (symbols).
For $\nu=1/3$, a noticeable difference is due to a slightly different magnetic field setting ($\delta B\simeq-0.5\,$T) between $V_\mathrm{gate}$ sweep and noise measurements.
} 
	\label{fig:Delta(tau)}
\end{figure*}

\vspace{\baselineskip}
{\noindent\textbf{Scaling dimension characterization.}}\\
\noindent
In previous characterizations of the charge $e^*$ of fractional quantum Hall quasiparticles, the shot noise is usually plotted as a function of the back-scattered current $I_\mathrm{B}$, and $e^*$ is extracted by matching the high-bias slope $\partial S/\partial I_\mathrm{B}$ with $2e^*(1-\tau)$, where $1-\tau$ corrects for tunneling correlations at finite $\tau$\cite{blanter_shotnoise_2000}. 
Even in this representation, which puts the emphasis on the larger high-bias shot noise, a visually discernible and experimentally relevant difference allows one to discriminate between predicted and trivial $\Delta$, as illustrated in Fig.~\ref{fig:TtoSNcrossover}a,b,c.
Blue continuous lines display the excess shot noise of quasiparticles of trivial $\Delta=1/2$ and of charge $e/3$ (a,c) or $e/5$ (b), which is given by the broadly used phenomenological expression\cite{blanter_shotnoise_2000,Heiblum_FracChargeMeas_2010}:
\begin{equation}
    S_{1/2}=2e^*I_\mathrm{B}(1-\tau)\left[\coth{\frac{e^*V}{2k_\mathrm{B}T}}-\frac{2k_\mathrm{B}T}{e^*V}\right].\label{eq:Sexc1s2}
\end{equation}
Note that for composite edges with several electrical channels (such as $\nu\in\{2/5,3\}$), $\tau\equiv I_\mathrm{B}/I_\mathrm{inj}$, $I_\mathrm{B}$ and $I_\mathrm{inj}$ refer to the dc transmission ratio and currents along the specific edge channel of interest (Methods).
The continuous lines of a different color in the main panels of Fig.~\ref{fig:TtoSNcrossover}a,b,c show the excess noise for the predicted quasiparticle scaling dimension $\Delta=1/6$ (red, (a)), $\Delta=3/10$ (purple, (b)) and $\Delta=1/3$ (green, (c)) obtained from\cite{Rosenow_NonUnivQPCfqhe_2004,Snizhko_ScaldimFano_2015,Schiller_ScaldimFano_2022}:
\begin{equation}
    S_{\Delta}=2e^*I_\mathrm{B}(1-\tau)\,\mathrm{Im}\left[\frac{2}{\pi}\psi\left(2\Delta+i\frac{e^*V}{2\pi k_\mathrm{B}T}\right)\right].\label{eq:SexcDelta}
\end{equation}
Here $\psi$ is the digamma function and $1-\tau$ the ad hoc amplitude factor used for extracting $e^*$ from the shot noise slope at high bias (beyond the perturbative limit $\tau\ll1$ where Eq.~\ref{eq:SexcDelta} is rigorously derived).
Whereas Eq.~\ref{eq:SexcDelta} reduces to Eq.~\ref{eq:Sexc1s2} for $\Delta=1/2$, for smaller $\Delta$ the shot noise emerges at a lower voltage. 
Intuitively, this can be connected through the time-energy relation to the slower decay of correlations at long times (as $t^{-2\Delta}$). 
For the quasiparticles $\{e/3,\Delta=1/6\}$ predicted at $\nu=1/3$, the apparent width of the crossover is more than twice narrower than for $\Delta=1/2$ (Fig.~\ref{fig:TtoSNcrossover}a).
The difference is smaller for the quasiparticles $\{e/5,\Delta=3/10\}$ and $\{e/3,\Delta=1/3\}$ since $\Delta$ is closer to $1/2$ (Fig.~\ref{fig:TtoSNcrossover}b,c).
Nevertheless, as can be straightforwardly inferred from the scatter of the data, it remains in all cases larger than our experimental resolution on the noise.
One can already notice that the illustrative shot noise measurements shown in Fig.~\ref{fig:TtoSNcrossover}a,b,c are closer to the parameter-free prediction of Eq.~\ref{eq:SexcDelta} with the expected $\Delta$. 
Note that this agreement is accompanied by a puzzling $I-V$ characteristics as previously mentioned (see $\tau(V)$ in insets and also in Extended Data Fig.~\ref{fig:SItau(V)}). 

For the present aim of characterizing $\Delta$ from the thermal-shot noise crossover, the Fano factor $F\equiv S/2eI_\mathrm{B}(1-\tau)$ of bounded amplitude at high bias is better suited\cite{Snizhko_ScaldimFano_2015,Schiller_ScaldimFano_2022}.
It is plotted versus the relevant variable $eV/k_\mathrm{B}T$ (see Eq.~\ref{eq:SexcDelta}) in Fig.~\ref{fig:TtoSNcrossover}d,e,f, with symbols and colored lines corresponding to the noise displayed in the panel immediately above.
Importantly, the effect of $\Delta<1/2$ on $F$ is not limited to an increased slope at low bias, which could in principle be attributed to a temperature lower than the separately characterized $T$, but results in marked changes in the overall shape of $F(eV/k_\mathrm{B}T)$.
In particular, for $\Delta=1/6$ the Fano factor is non-monotonous (red line in Fig.~\ref{fig:TtoSNcrossover}d).
The increasing steepness while reducing $\Delta$ combined with an overall change of shape facilitates the extraction of this parameter from a fit using Eq.~\ref{eq:SexcDelta}.
Qualitatively, the value of $F$ at large bias solely reflects $e^*/e$, the overall crossover shape (such as a non-monotonous dependence at $\Delta<1/4$) solely involves $\Delta$, and the low-bias slope is a combination of both $e^*$ and $\Delta$.
The results of such fits (minimizing the data-Eq.~\ref{eq:SexcDelta} variance) are shown as black continuous lines in Fig.~\ref{fig:TtoSNcrossover}d,e,f, together with the corresponding fitting parameters $e^*$ and $\Delta$ (the temperature being fixed to the separately determined $T\simeq31$\,mK).

\vspace{\baselineskip}
{\noindent\textbf{Robustness of observations.}}\\
\noindent
Focusing on the Fano factor cancels out some of the non-universal behaviors, but not all of them.
Of particular concern are the disorder-induced resonances, which could result in a Coulomb-dominated sequential tunneling with a strong effect on the Fano factor.
This is likely to happen in the fractional quantum Hall regime where QPCs often exhibit narrow peaks and dips in their transmission $\tau$ versus gate voltage (see insets in Fig.~\ref{fig:Delta(tau)}).
Accordingly, for some gate voltages we find that an accurate fit of the noise data is not possible with Eq.~\ref{eq:SexcDelta}, whatever $e^*$ and $\Delta$.
In such cases, the resulting fitted values are meaningless.
This was transparently addressed with a maximum variance criteria between data and best fit.
If the fit-data variance is higher, the extracted $e^*$ and $\Delta$ are discarded (see Methods).
This same procedure was systematically applied to all the noise measurements performed over a broad span of gate voltages controlling $\tau$ (the full data set, including discarded fits and analysis code, is available in a Zenodo deposit).

The values of $e^*$ and $\Delta$ obtained while spanning the gate voltage of the same QPC$_\mathrm{E}$ are shown versus $\tau(V=0)$ in Fig.~\ref{fig:Delta(tau)}, for each of the three probed fractional quasiparticles (see Methods for electrons at $\nu=3$).
We find remarkably robust scaling dimensions (and charges), close to the predictions shown as horizontal lines.
In particular, although the nature of the tunneling quasiparticles is eventually going to change at $\tau\rightarrow1$, we observe that $\Delta$ and $e^*$ extracted with Eq.~\ref{eq:SexcDelta} (which is exact only at $\tau\ll1$) remain relatively stable over a broad range of $\tau$. 
Such a stability matches previous $e^*$ measurements, including a particularly steady $e/5$\cite{Griffiths_EstarVsTau_2000}. 
Figure~\ref{fig:Delta(tau)}a shows data points obtained in the $\nu=1/3$ plateau.
A statistical analysis of the ensemble of these points yields $\braket{\Delta}\simeq0.167$ with a spread of $\sigma_\Delta\simeq0.023$, which is to be compared with the prediction $\Delta=1/6\simeq0.1667$.
The data-prediction agreement on $\Delta$ is at the level, if not better, than that on $e^*$ (often found slightly lower than expected).
Similar sweeps are shown in Fig.~\ref{fig:Delta(tau)}b,c for the inner channel of conductance $e^2/15h$ at $\nu=2/5$ (b), and at $\nu=2/3$ (c).
Note that a few data points at $\nu=2/5$ and at $\nu=2/3$ are displayed as pairs of `$\times$' ($e^*/e$) and `$+$' ($\Delta$) instead of open and full symbols (Fig.~\ref{fig:Delta(tau)}, panels b and c).
This indicates an anomalous fitted value of the charge $e^*$, off by about 50\% or more from the well-established prediction $e^*_\mathrm{th}=e/5$ and $e^*_\mathrm{th}=e/3$ respectively (dash-dot line).
Because this suggests a non-ideal QPC behavior (e.g.\ involving localized electronic levels), we chose not to include these relatively rare points in the data ensemble analysis of $\Delta$ (they remain included in the statistical analysis of $e^*$).
For this reduced data set composed of fifteen measurements along the inner channel at $\nu=2/5$, we obtain $\braket{\Delta}\simeq0.327$ $(\sigma_\Delta\simeq0.078)$, which is to be compared with the predicted $\Delta=3/10$ of $e/5$ quasiparticles.
Lastly, at $\nu=2/3$, the gate voltage sweep shown in Fig.~\ref{fig:Delta(tau)}c gives $\braket{\Delta}\simeq0.249$ ($\sigma_\Delta\simeq0.029$), close to the predicted $\Delta=1/3\simeq0.33$. 
Note however that in this more complex case, with counter-propagating edge modes and the emergence of a small plateau vs gate voltage at $\tau\simeq0.5$ (inset in Fig.~\ref{fig:Delta(tau)}c), the noise interpretation is not as straightforward, especially when $\tau$ is not small (see Methods for additional tests and discussions).

\begin{figure}[bh!]
	\centering
	\includegraphics[width=9cm]{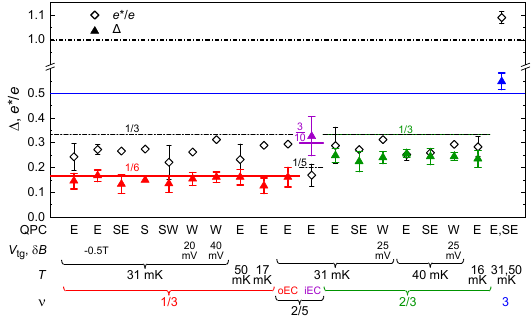}
	\caption{\textbf{Summary of observations.}
 Each symbol with error bars represents the mean and standard deviation of the ensemble of $\Delta$ ($\blacktriangle$) and $e^*/e$ ($\diamond$) extracted along one gate voltage span of a QPC, such as those shown in Fig.~\ref{fig:Delta(tau)}.
 The horizontal axis indicates the experimental conditions: label of the QPC (E, SE, S, SW, W), voltage $V_\mathrm{tg}$ applied to the top gate (TG) around QPC$_\mathrm{W}$, magnetic field shift $\delta B$ from the center of the plateau (if any), temperature $T$, filling factor $\nu$ and for $\nu=2/5$ the probed edge channel (inner `iEC' or outer `oEC') with different colors for different predicted quasiparticles.
 }
	\label{fig:summary}
\end{figure}

The robustness and generic character of these $\Delta$ observations are further established by repeating the same procedure in different configurations:
\textit{(i)} on several QPCs, with different orientations with respect to the Ga(Al)As crystal; \textit{(ii)} for multiple temperatures $T$; \textit{(iii)} for several top gate voltages $V_\mathrm{tg}$ controlling the density in the vicinity of QPC$_\mathrm{W}$; \textit{(iv)} by changing the magnetic field, both along the $\nu=1/3$ plateau and to $\nu=2/5$ on the outer edge channel.
Figure~\ref{fig:summary} recapitulates all our measurements (283 in total), including conventional electrons at $\nu=3$. 
Each point represents the average value $\braket{e^*/e}$ ($\diamond$) or $\braket{\Delta}$ ($\blacktriangle$) and the corresponding standard deviation obtained while broadly spanning the gate voltage of the indicated QPC (individually extracted $e^*$ and $\Delta$ are provided in Methods).
See also Methods for consistent conclusions from an alternative fitting procedure where $\Delta$ is the only free parameter ($e^*$ being fixed to the well-established prediction, and focusing on low voltages $e^*|V|\leq 2k_\mathrm{B}T$).

\vspace{\baselineskip}
{\noindent\textbf{Conclusion.}}\\
\noindent
Fano factor measurements previously used to investigate the charge of tunneling quasiparticles also allow for a consistent determination of their scaling dimension, from the width and specific shape of $F(eV/k_\mathrm{B}T)$.
Combined with a systematic approach, the resulting observations of $\Delta$ establish long-lasting theoretical predictions for the fractional quantum Hall quasiparticles at $\nu=1/3$, $2/5$ and $2/3$. 
This approach could be generalized to other quasiparticles, with the potential to shed light on the non-abelian quasiparticles predicted at even-denominator filling factors. 
It may also be applied to other low-dimensional conductors.


\vspace{\baselineskip}
\small

{\noindent\textbf{Data and code availability.}}
Plotted data, raw data and data analysis code are available
on Zenodo: https://doi.org/10.5281/zenodo.10599318

{\noindent\textbf{Acknowledgments.}}
This work was supported by the European Research Council (ERC-2020-SyG-951451) and the French RENATECH network.
We thank K.~Snizhko for discussions and E.~Boulat for providing the $\tau(V,T)$ prediction at $\nu=1/3$ in Fig.~\ref{fig:TtoSNcrossover}.

{\noindent\textbf{Author Contributions.}}
A.V., C.P., P.G.,  Y.S. and F.P. performed the experiments with inputs from A.Aa. and A.An.;
A.V. and F.P. analyzed the data with inputs from A.An., C.P., P.G. and Y.S.;
A.C. and U.G. grew the 2DEG;
A.V., A.Aa, and F.P. fabricated the sample;
Y.J. fabricated the HEMT used in the cryogenic noise amplifiers;
A.V. and F.P. wrote the manuscript with contributions from all authors;
A.An. and F.P. led the project.

{\noindent\textbf{Author Information.}}
Correspondence and requests for materials should be addressed to A.An. (anne.anthore@c2n.upsaclay.fr) and F.P. (frederic.pierre@cnrs.fr).\\

\noindent\textit{Note added.}-- Coincident to the present investigation, two other works are experimentally addressing the scaling dimension of the $e/3$ fractional quantum Hall quasiparticles at $\nu=1/3$.
An experiment by the team of M.~Heiblum with a theoretical analysis led by K.~Snizhko (N.~Schiller \textit{et al.}, arXiv:2403.17097) exploits the same thermal to shot noise crossover as in the present work, with a focus on low voltages and assuming the predicted fractional charge (see Fig.~\ref{fig:summaryRounding} for such a single parameter data analysis at low bias), and finds $\Delta\simeq1/2$.
The team of G.~Feve (M.~Ruelle \textit{et al.}, arXiv:2409.08685) relies on a different, dynamical response signature and finds $\Delta\simeq1/3$.
In these two coincident works the extracted scaling dimension is different from the pristine prediction $\Delta=1/6$, which is observed in the present work.
As pointed out in the manuscript, the emergence of non-universal behaviors could be related to differences in the geometry of the QPCs.

\normalsize
\vspace{\baselineskip}
{\Large\noindent\textbf{METHODS}}
\small

\begin{figure*}[!htb]
\renewcommand{\figurename}{\textbf{Extended Data Figure}}
	\centering
	\includegraphics[width=18cm]{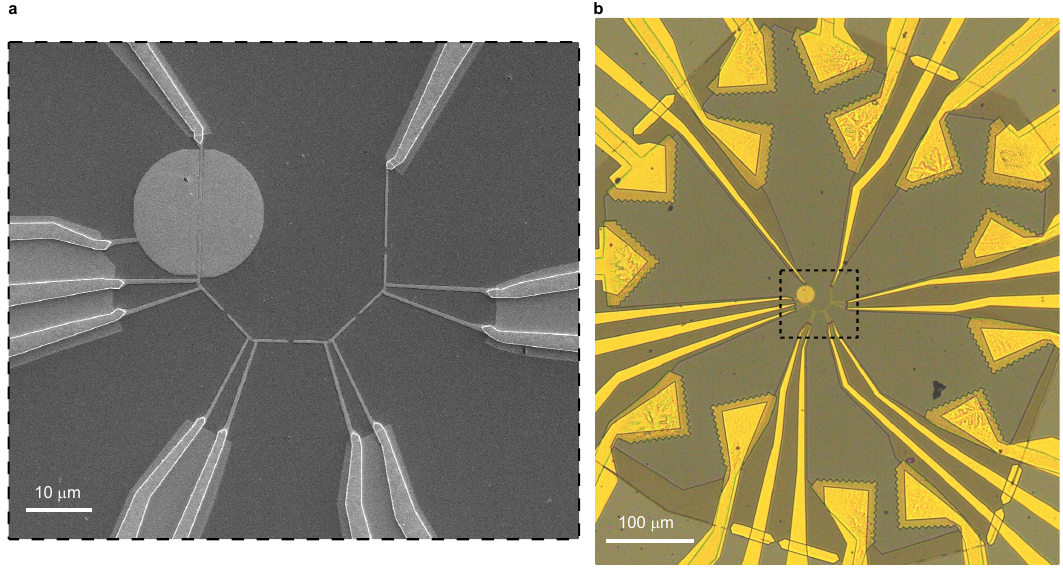}
	\caption{
	\footnotesize
	\textbf{Large scale pictures of measured device. a}, electron-beam micrograph. 
 Areas with a 2DEG underneath (the mesa) appear darker.
 Lighter parts with bright edges are thick gold layers used to climb down the mesa edges and as bonding pads. 
 \textbf{b}, optical image. 
 The thick top layer made of gold appears as the brightest yellow.
 Ohmic contacts have staircase edges and show as a darker shade of yellow.
 The surface over which the HfO$_2$ was deposited (dark grey) completely encapsulates the active part of the device, including ohmic contacts. 
 The dashed rectangle indicates the boundary of the electron-beam image in panel a.
	}	\label{fig:SIsample}
\end{figure*}

\begin{figure*}[h]
\renewcommand{\figurename}{\textbf{Extended Data Figure}}
	\centering
	\includegraphics[width=14cm]{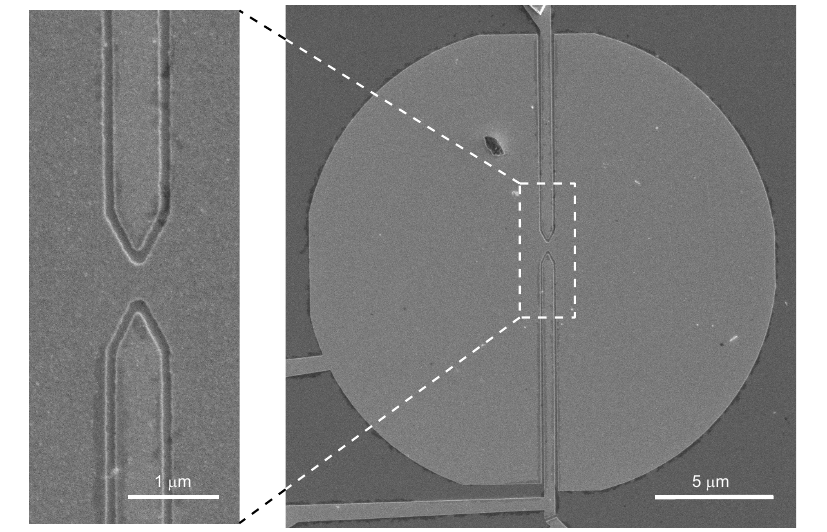}
	\caption{
	\footnotesize
	\textbf{Quantum point contact with surrounding metal gate.} Electron-beam micrographs of QPC$_\mathrm{W}$.
	}	\label{fig:SIQPCW}
\end{figure*}

{\noindent\textbf{Sample.}} 
The sample is nanofabricated by e-beam lithography on a Ga(Al)As heterojunction forming a 2DEG buried at 140\,nm, of density $n=1.2\times10^{11}$\,cm$^{-2}$, and of mobility $1.8\times10^6$\,cm$^2$V$^{-1}$s$^{-1}$.
The 2DEG mesa is first delimited by a wet etching of 105\,nm, deeper than the Si $\delta$-doping located 65\,nm below the Ga(Al)As surface.
The large ohmic contacts (schematically displayed as circles in Fig.~\ref{fig:sample}) used to drive and measure the quantum Hall edge currents are then formed $100-200\,\mu$m away from the QPCs by e-beam evaporation of a AuGeNi stack followed by a 50s thermal annealing at 440$^\circ$C.
A 15\,nm layer of HfO$_2$ is grown by thermal ALD at 100$^\circ$C over the entire mesa, in order to strongly reduce a gate-induced degradation of the 2DEG that could complicate the edge physics.
This degradation is generally attributed to unequal thermal contractions upon cooling\cite{Davis_GateGaAsStrain_1994} or a deposition stress, which could also modulate the edge potential carrying the quantum Hall channels along the gates.
In previous works, we observed a change in the behavior of QPCs, including in the thermal to shot noise crossover, that was correlated with their orientation\cite{Glidic_Collider2023,Glidic_Andreev_2023} (see also source vs central QPCs in Refs.~\citenum{Bartolomei_Cross_1tiers_2020,Ruelle_2s5_2023}), which we suspect to result from such gate induced complication.
Here the Ti (5\,nm) - Au (20\,nm) gates used to form the QPCs are evaporated on top of the HfO$_2$.
The five QPCs, of different orientations with respect to the Ga(Al)As crystal, have nominally identical geometries.
The split gates have a nominal tip-to-tip distance of 600\,nm and a 25$^{\circ}$ tip opening angle prolonged until a gate width of 430\,nm.
Larger-scale electron-beam and optical images of the measured device are displayed in Extended Data Figure~\ref{fig:SIsample}.
The relatively important gate width (about three times the 2DEG depth) was chosen to reduce possible complications from Coulomb interactions between the quantum Hall edges across the gates\cite{Papa_anomTDOSinteraction_2004,Kamata_artificialTLL_2014}, and to better localize the tunneling location when the QPC is almost open (for less negative gate voltages)\cite{Dolcetto_TDOSextendedQPC_2012}.
The nominal separation between the split gates controlling QPC$_\mathrm{W}$ and the surrounding metal gate is 150\,nm.
A high magnification picture of QPC$_\mathrm{W}$ with its surrounding metal gate is shown in Extended Data Figure~\ref{fig:SIQPCW}.
Note that all the gates were grounded during the cooldown.\\

{\noindent\textbf{Measurement.}} 
The sample is cooled in a cryofree dilution refrigerator and electrically connected through measurement lines both highly-filtered and strongly anchored thermally (see Ref.~\citenum{Iftikhar_Ttriad_2016} for details).
Final $RC$ filters with CMS components are positioned within the same metallic enclosure screwed to the mixing chamber that holds the sample: 200\,k$\Omega$-100\,nF for gate lines, 10\,k$\Omega$-100\,nF for the bias line, and 10\,k$\Omega$-1\,nF for low frequency measurement lines.
Note a relatively important filtering of the bias line, which prevents an artificial rounding of the thermal noise-shot noise crossover from the flux noise induced by vibrations in a magnetic field.
The differential QPC transmission $\partial I_\mathrm{B}/\partial I_\mathrm{inj}=1-\partial I_\mathrm{F}/\partial I_\mathrm{inj}$ is measured by standard lock-in techniques at 13\,Hz.
A particularly small ac modulation is applied on $V$ (of rms amplitude $V_\mathrm{ac}^\mathrm{rms}\approx k_\mathrm{B}T/3e$) to avoid any discernible rounding of the thermal noise-shot noise crossover.
The transmitted and reflected dc currents used to calculate $\tau$ and $F$ are obtained by integrating with the applied bias voltage the corresponding lock-in signal $I_\mathrm{B,F}(V)=\int_0^V(\partial I_\mathrm{B,F}/\partial V)dV$.

A specific QPC is individually addressed by completely closing all the other ones.
For the composite edges at $\nu=2/5$ and $\nu=3$, the characterizing current transmission ratio $\tau$ refers to the current transmission along the specific channel of interest. 
Explicitly, at $\nu=2/5$, the transmission $\tau$ along the inner edge channel is given by the ratio between measured (total) $I_\mathrm{B}^\mathrm{meas}$ (only the inner channel of interest is back-scattered, the outer channel is fully transmitted as attested by a broad and noiseless $e^2/3h$ plateau) normalized by the current $Ve^2/15h$ injected along the inner channel: $\tau=I_\mathrm{B}^\mathrm{meas}/(Ve^2/15h)$.
For the outer channel at $\nu=2/5$, the fully back-scattered inner edge channel current $Ve^2/15h$ is removed from the measured total $I_\mathrm{B}^\mathrm{meas}$ and the result is normalized by the current $Ve^2/3h$ injected along the outer edge channel: $\tau=(I_\mathrm{B}^\mathrm{meas}-Ve^2/15h)/(Ve^2/3h)$.

Noise measurements are performed using specific cryogenics amplification chains connected to dedicated ohmic contacts, through nearly identical $L-C$ tanks of resonant frequency $0.86\,$MHz\cite{Liang_HEMTs_2012,Jezouin_QLimHeatFlow_2013b}.
The noise ohmic contacts are located upstream of the ohmic contacts used for low frequency transmission measurements, as shown in Fig.~\ref{fig:sample}.
A dc block (2.2\,nF) at the input of the $L-C$ tanks preserves the low frequency lock-in signal.
For the particular case of QPC$_\mathrm{E}$, the forward (transmitted) current fluctuations $\delta I_\mathrm{F}$ are also measured, which gives us access to $\braket{\delta I_\mathrm{F}^2}$ and to the cross-correlations $\braket{\delta I_\mathrm{B}\delta I_\mathrm{F}}$.
Besides increasing the signal to noise ratio for QPC$_\mathrm{E}$, this allows one to ascertain that $\braket{\delta I_\mathrm{B}^2}$ matches the more robust cross-correlation signal\cite{Barta_AutovsCross_2023}.

\begin{figure*}[h]
\renewcommand{\figurename}{\textbf{Extended Data Figure}}
	\centering
	\includegraphics[width=9cm]{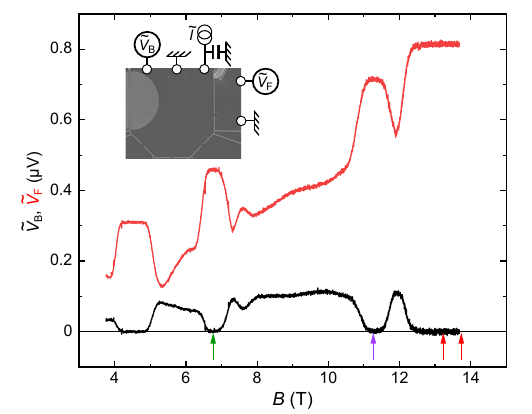}
	\caption{
	\footnotesize
	\textbf{Magnetic field sweep.} 
Forward ($\Tilde{V}_\mathrm{F}$, red) and back-scattered ($\Tilde{V}_\mathrm{B}$, black) ac voltages across a fully open QPC$_\mathrm{E}$, in response to a fixed ac bias current $\Tilde{I}$, are plotted as a function of magnetic field $B$.
The other QPCs are fully closed.
Arrows indicate at which $B$ the different measurements were performed (except $B\simeq1.5\,$T at $\nu=3$, not shown here).
At these points the back-scattered signal is zero whereas the forward signal is well within a plateau, despite the increased mixing chamber temperature of 160\,mK during this $B$ sweep.
Note that $\Tilde{V}_\mathrm{F}$ does not precisely scale as $h/\nu e^2$ along plateaus, due to the parallel capacitance shown schematically (100\,nF) and the finite ac frequency (13\,Hz).
}	\label{fig:SIVfb(B)}
\end{figure*}

\begin{figure}[tbh]
\renewcommand{\figurename}{\textbf{Extended Data Figure}}
	\centering
	\includegraphics[width=9cm]{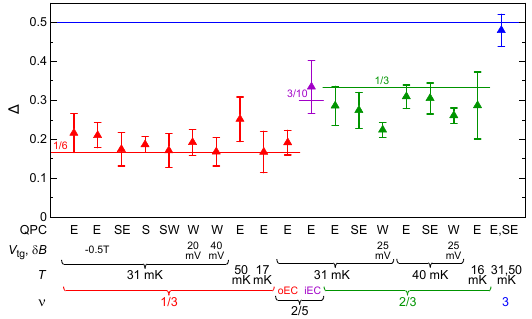}
	\caption{\textbf{Summary of single parameter analysis.}
  Symbols recapitulate the extracted scaling dimension in all explored configurations, similarly to Fig.~\ref{fig:summary} but with $\Delta$ obtained by a different procedure.
  The quasiparticule charge $e^*$ is here assumed to take its predicted value, and the fit is performed solely on the thermal-shot noise crossover at low bias and using $\Delta$ as only free parameter (see text).
  }
	\label{fig:summaryRounding}
\end{figure}

The device was immersed in a magnetic field close to the center of the corresponding Hall resistance plateaus, except when a shift $\delta B$ is specifically indicated.
The data at $\nu=1/3$, $\nu=2/5$, $\nu=2/3$ and $\nu=3$ were obtained at $B=13.7\,$T (13.2\,T for $\delta B=-0.5\,$T), 11.3\,T, 6.8\,T and $1.5$\,T, respectively.
See vertical arrows in Extended Data Fig.~\ref{fig:SIVfb(B)} for the localization of these working points within a magnetic field sweep of the device along $B\in[4,14]$\,T ($\nu\in[1/3,1]$).\\

{\noindent\textbf{Thermometry.}} 
The electronic temperature inside the device is ascertained by the noise measured at thermal equilibrium, with all QPCs closed.
For temperatures $T\geq30\,$mK (up to the maximum $T\simeq55$\,mK), we find at $\nu=1/3$ and $\nu=3$ that the measured thermal noise is linear with the temperature readings of our calibrated RuO$_2$ thermometer. 
This establishes the good thermalization of electrons in the device with the mixing chamber, as well as the calibration of the RuO$_2$ thermometer.
Accordingly, we indifferently get $T\geq30\,$mK from the equilibrium noise or the equivalent RuO$_2$ readings.
At the lowest probed temperatures $\sim15$\,mK, the RuO$_2$ thermometers are no longer reliable and $T$ is obtained from the thermal noise, by linearly extrapolating from $S(T\geq30\,$mK$)$.
Note that the $S(T)$ slope was not recalibrated at $\nu=2/3$, but its change from $\nu=1/3$ was calculated from the separately obtained knowledge of the $L-C$ tank circuit parameters, see `Noise amplification chains calibration' below.\\

{\noindent\textbf{Noise amplification chains calibration.}} 
The gain factors $G^\mathrm{eff}_\mathrm{F,B}$ between raw measurements of the auto-correlations, integrated within a frequency range $[f_\mathrm{min},f_\mathrm{max}]$, and the power spectral density of current fluctuations $\braket{\delta I_\mathrm{F,B}^2}$ are obtained from:
\begin{equation}
    G_\mathrm{F,B}^\mathrm{eff}=\frac{s_\mathrm{F,B}}{4k_\mathrm{B}(1/R_\mathrm{tk}+\nu e^2/h)},
    \label{eq:gaincalib}
\end{equation}
with $R_\mathrm{tk}\simeq150\,\mathrm{k}\Omega$ the effective parallel resistance accounting for the dissipation in the considered $L-C$ tank, and $s_\mathrm{F,B}$ the slope of the raw thermal noise versus temperature.
The cross-correlation gain factor is simply given by $G_\mathrm{FB}^\mathrm{eff}\simeq \sqrt{G_\mathrm{F}^\mathrm{eff}G_\mathrm{B}^\mathrm{eff}}$, up to a negligible reduction ($<0.5\%$) due to the small difference between the two $L-C$ tanks.
In practice, the thermal noise slopes $s_\mathrm{F,B}$ where only measured at $\nu=1/3$ and $\nu=3$.
The changes in $G^\mathrm{eff}_\mathrm{F,B}(\nu)$ at $\nu\in\{2/3,2/5\}$ from the gains at $\nu=1/3$ are obtained from:
\begin{equation}
    \frac{G_\mathrm{F,B}^\mathrm{eff}(\nu)}{G_\mathrm{F,B}^\mathrm{eff}(1/3)}=\frac{\int_{f_\mathrm{min}}^{f_\mathrm{max}}|Z_\mathrm{tk}^{-1}(f)+\nu e^2/h|^{-2}df}{\int_{f_\mathrm{min}}^{f_\mathrm{max}}|Z_\mathrm{tk}^{-1}(f)+e^2/3h|^{-2}df},
    \label{eq:gainchange}
\end{equation}
with the tank impedance given by $Z_\mathrm{tk}^{-1}(f)=R_\mathrm{tk}^{-1}+(i L_\mathrm{tk}2\pi f)^{-1} + i C_\mathrm{tk}2\pi f$, where $L_\mathrm{tk}\simeq250\,\mu$H and $C_\mathrm{tk}\simeq135$\,pF (see Methods in Ref.~\citenum{Glidic_Andreev_2023} for details regarding the calibration of the tank parameters).
At $\nu\in\{2/3,2/5,1/3\}$, we integrated the noise signal in the same frequency window $f_\mathrm{min}=840\,$kHz and $f_\mathrm{max}=880$\,kHz.
At $\nu=3$, a larger window $f_\mathrm{min}=800\,$kHz and $f_\mathrm{max}=920$\,kHz takes advantage of the larger bandwidth $\sim\nu e^2/2\pi h C_\mathrm{tk}$.\\

{\noindent\textbf{Noise tests.}} 
Among various experimental checks, we point out: 
\textit{(i)} The effect of a dc bias voltage on the noise when each of the QPCs are either fully open or fully closed, which is here found below our experimental resolution.
The presently imperceptible `source' noise could have resulted from poor ohmic contact quality, incomplete electron thermalization in the contacts, dc current heating of the resistive parts of the bias line...
\textit{(ii)} The effect of the QPC transmission on the noise at zero dc bias voltage, which is here negligible at our experimental resolution. 
This rules out a possibly higher electron temperature in the ohmic contact connected to the bias line with respect to one connected to a cold ground, which would translate into an increase in $\braket{\delta I_\mathrm{B}^2}$ at $\tau=1$ compared to $\tau=0$.
It also shows that the vibration noise in the bias line at frequencies well below 1\,MHz does not translate into a discernible broadband excess shot noise for intermediate values of $\tau$.\\

{\noindent\textbf{Fitting details.}} 
The extracted values of $e^*$ and $\Delta$ shown in Fig.~\ref{fig:Delta(tau)} and in Extended Data Figs.~\ref{fig:SIDelta(tau)_1s31s6},\ref{fig:SIDelta(tau)_2s33} represent the best fit parameters minimizing the variance between the shot noise data and Eq.~\ref{eq:SexcDelta}. 
Only the meaningful points are displayed and included in the statistical analysis.
These fulfill two conditions: \textit{(i)} an accurate fit of the data can be achieved and \textit{(ii)} the charge does not deviate too much from the expected value. 
Condition \textit{(i)} requires a quantitative assessment of the fit accuracy.
We used for this purpose the coefficient of determination $R^2$ and chose to apply the same threshold to all the data taken in similar conditions.
Specifically, we automatically discarded fits of $R^2<0.9965$ at $\nu=1/3$ and for the outer channel at $\nu=2/5$, $R^2<0.9966$ for the inner channel at $\nu=2/5$, and $R^2<0.9968$ at $\nu=2/3$.
The number of $S(V)$ sweeps discarded by condition \textit{(i)} is important, 2/3 of the total number (mostly when $\tau$ is too close to 0 or 1). 
We checked that the overall results are only marginally affected by the specific threshold value (within reasonable variations).
All the points that satisfy condition \textit{(i)} are displayed and included in the statistical analysis of the quasiparticle charge.
Condition \textit{(ii)} is subsequently applied to deal with situations where the fitting charge is found at odds with the predicted value. Specifically we discarded $S(V)$ sweeps for which the charge is found to be more than 44\% off, i.e. $e^*<0.56e^*_\mathrm{th}$ or  $e^*>1.44e^*_\mathrm{th}$. 
The former happens at small $\tau$ with a small QPC gate voltage. This gate voltage might not be enough to deplete the gas under the QPC gates, which could make tunneling happen in several places along the gates, and not only located at their tip, deviating from the model of a point contact.  
The latter occurs in the so-called strong back-scattering regime where the nature of the tunneling quasiparticles is expected to change. Indeed, in the weak back-scattering regime ($\tau\ll 1$) the tunnneling barrier between the two edges is made of the electron gas in the fractional quantum Hall regime that selects the quasiparticles. However in the strong back-scattering regime ($\tau\rightarrow1$) the tunnneling barrier between the two edges is made of vacuum that will select electrons.
The points that do not satisfy condition \textit{(ii)} are displayed with different symbols and not included in the statistical analysis of $\Delta$. 
They represent a small fraction (5\%) of the data satisfying \textit{(i)}.

A complementary fitting procedure was employed to further establish the robustness of our results.
In Extended Data Fig.~\ref{fig:summaryRounding} we summarize the extracted $\Delta$ obtained by fitting the thermal to shot noise crossover of $S(V)$ using for $e^*$ the theoretically predicted value.
The fits are performed on the same set of $S(V)$ sweeps as for the main fitting procedure for $\Delta$ (obeying the two above mentioned conditions \textit{(i)} and \textit{(ii)}).
The voltage bias extension of these fits is reduced to $e^*|V|\leq2k_\mathrm{B}T$ to limit the weight of the shot noise solely sensitive to $e^*$.\\

{\noindent\textbf{Predictions.}} 
The $\Delta$ predictions indicated in the manuscript for fractional quasiparticles at $\nu=1/3$ and $2/5$ follow the Luttinger liquid expression $\Delta=(e^*/e)^2/(2Gh/e^2)$ for $e^*$ quasiparticles along a chiral 1D channel of conductance $G$\cite{Wen_BookQFT_2004,Kane_ImpScattFQHE_1995,Schiller_ScaldimFano_2022}.
Note that in these fully chiral states (where all channels along one edge propagate in the same direction), the quasiparticles exchange phase $\theta$ is predicted to be simply related to $\Delta$ by the relation $\theta=2\pi\Delta$ (see e.g.\ Appendix~A in Ref.~\citenum{Schiller_ScaldimFano_2022}).

The above Luttinger expression for $\Delta$ does not apply at $\nu=2/3$ for $e/3$ quasiparticles delocalized between a $2e^2/3h$ edge channel, and a neutral counter-propagating channel that further increases $\Delta$, see Ref.~\citenum{Kane_2s3disorder_1994}.
Note that, in general, the predicted link between $\Delta$ and $\theta$ for fully chiral quantum Hall edges does not hold in the presence of counter-propagating (charged and/or neutral) modes\cite{Schiller_ScaldimFano_2022}.\\

{\noindent\textbf{Filling factor 2/3.}} 
In this more complex hole conjugate state\cite{Wen_BookQFT_2004,Jain_BookCompositeFermions_2007,Kane_2s3disorder_1994} \textit{(i)} the edges are not fully chiral and found to carry a backward heat current (going upstream the flow of electricity), and \textit{(ii)} the QPCs can exhibit a plateau at half transmission (see e.g.\ Refs.~\citenum{Gross_UpstreamNeutral2s3,Bid_2s3noisyplateau_2009}).
The former may introduce unwanted heat-induced contributions to the noise while the latter alludes to a composite edge structure. 
Both have possible consequences on the interpretation of the noise signal\cite{Mirlin_NoisyPlateaus_2020,Gefen_NoisyPlateaus_2021,Goldstein_NoisyPlateaus_2023}.

\begin{figure*}[tbh]
\renewcommand{\figurename}{\textbf{Extended Data Figure}}
	\centering
	\includegraphics[width=18cm]{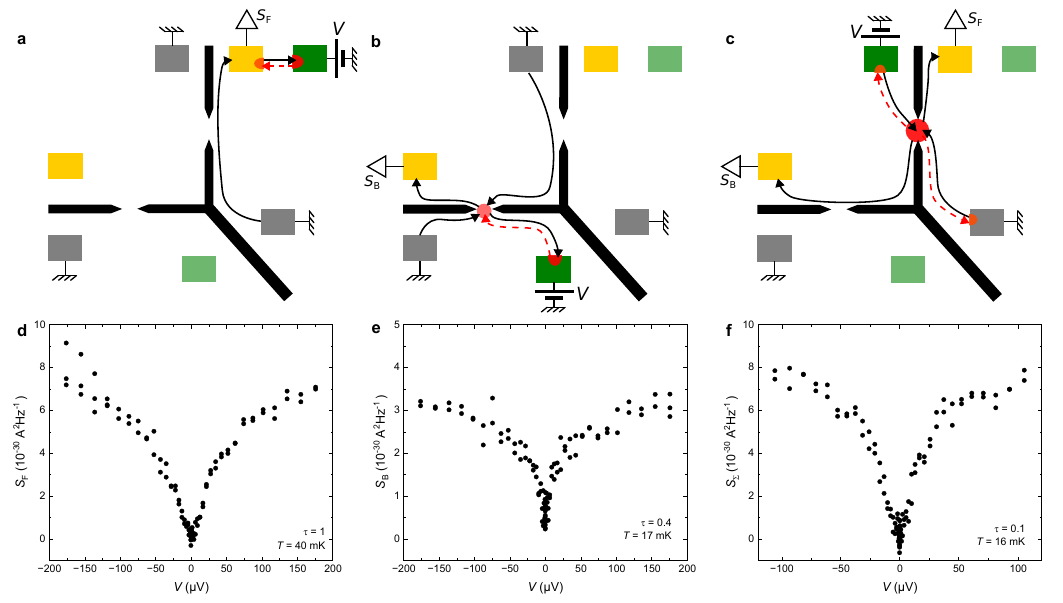}
	\caption{\textbf{Signatures of upstream neutral heat flow at $\nu=2/3$.}
  The presence of a neutral heat current flowing in the opposite direction of the electrical current is assessed by noise measurements in three different configurations. 
  \textbf{a, b, c,} Schematic illustrations of the three processes (described in the text) by which heat is created, transported upstream by the neutral modes (red dashed arrows) and detected. 
  \textbf{d, e, f,} Noise signature of upstream heating measured in the configuration displayed in the panel immediately above. 
  }
	\label{fig:Heating}
\end{figure*}

\textit{(i) Non-chiral heating.} As in previous works (see e.g.\ Ref.~\citenum{Gross_UpstreamNeutral2s3}), we observe in our device an upstream heating (only) at $\nu=2/3$, through three noise signatures (see Extended Data Fig.~\ref{fig:Heating}).
(Signature~1)~The strongest noise signature, seen at all temperatures (see Extended Data Fig.~\ref{fig:Heating}d for $T\simeq40$\,mK), is obtained in the configuration schematically depicted Extended Data Fig.~\ref{fig:Heating}a. 
Here the noise is measured on a contact located electrically upstream a hot spot in an adjacent voltage biased contact ($\sim30\,\mu$m away, e.g.\ the contact normally used for measuring $\braket{I_\mathrm{B}}$ in Fig.~\ref{fig:sample}). 
As illustrated in Extended Data Fig.~\ref{fig:Heating}a, the noise increase is attributed to a local heating of the noise measurement contact (near the location where electrical current is emitted from this contact) by the upstream neutral heat current originating from the downstream hot spot.
See Fig.~3 of Ref.~\citenum{Gross_UpstreamNeutral2s3} for a similar observation in the same configuration. 
Note that in configurations used for investigating $\Delta$, the heat generated at the downstream grounded contacts cannot propagate to the noise measurement contacts, since a floating contact located in between (measuring $\braket{I_\mathrm{B,F}}$, see Fig.~\ref{fig:sample}) absorbs the upstream heat flow (see Appendix~A and Fig.~10 in Ref.~\citenum{Mirlin_NoisyPlateaus_2020} for a specific discussion). 
(Signature~2)~A weaker noise signature from a different heating mechanism is observed, only at the lowest temperature ($T\simeq17$\,mK in Extended Data Fig.~\ref{fig:Heating}e), in the configuration schematically depicted Extended Data Fig.~\ref{fig:Heating}b (the same configuration is labeled N$\rightarrow$C in Fig.~4 of Ref.~\citenum{Gross_UpstreamNeutral2s3}).
Here a hot spot is created at a downstream contact biased at $V$.
The counter propagating neutral mode carries an upstream heat current to the QPC, which converts the increased temperature into electrical noise from a thermally induced mechanism.
The signal is weaker, as would be expected from a smaller heat current through the longer distance of $\sim150\,\mu$m between hot spot and QPC (the heat propagation is diffusive due to thermal equilibration between counter-propagating channels).
In practice, it is discernible only at the lowest temperature $T\simeq17\,$mK and for the highest QPC sensitivity ($\tau\sim0.5$).
The lower effect (imperceptible here) at higher temperatures is expected from the generally more efficient relaxation to thermal equilibrium.
Note that in the configurations used to probe $\Delta$, the contacts immediately downstream the QPC are floating (used to measure the noise or $\braket{I_\mathrm{F}}$) and, consequently, absorb the upstream heat current originating from the grounded contacts further downstream.
(Signature~3) A possibly more consequential signature of upstream heating is observed in the same configuration as that used to probe $\Delta$, through an increase in the noise sum $S_\Sigma\equiv S_\mathrm{F}+S_\mathrm{B}+2S_\mathrm{FB}$. 
From charge conservation and the chirality of electrical current, $S_\Sigma$ corresponds to the thermal noise emitted from the source contacts electrically upstream the QPC (independently of any noise generated along the path, such as the partition noise at the QPC, as long as there is no charge accumulation in the probed MHz range).
In fully chiral states such as $\nu\in\{1/3,2/5,3\}$ the source contacts' temperature is independent of the applied bias $V$ (at the emission point), and so is $S_\Sigma$.
At $\nu=2/3$ and $T\simeq16\,$mK this is not the case, as shown in Extended Data Fig.~\ref{fig:Heating}f.
This increase in $S_\Sigma$ is interpreted as the signature of a local hot spot in the $\sim 150\,\mu$m upstream source contacts, by heated-up counter-propagating neutral modes generated at the voltage-biased QPC as schematically illustrated Extended Data Fig.~\ref{fig:Heating}c (for a previous observation of the same mechanism, see configuration labeled C$\rightarrow$N in Fig.~4 of Ref.~\citenum{Gross_UpstreamNeutral2s3}). 
In practice, we observe a fast increase followed by a near saturation at $S_\Sigma\lesssim7\,10^{-30}\,\mathrm{A}^2\mathrm{Hz}^{-1}$, which is not negligible with respect to the partition noise of interest (see Fig.~\ref{fig:TtoSNcrossover}c).
To limit the impact of this effect at $T\simeq16\,$mK and $\nu=2/3$, we only considered the cross-correlation signal ($S\equiv-S_{FB}=-\braket{\delta I_\mathrm{F}\delta I_\mathrm{B}}$) measured on QPC$_\mathrm{E}$.
Indeed, a symmetric heating of the two source contacts electrically upstream of QPC$_\mathrm{E}$ (biased at $V$ and grounded) would not result in any change of the cross-correlations (but in a thermally induced increase of the auto-correlations).
See Ref.~\citenum{Barta_AutovsCross_2023} for a discussion regarding the stronger robustness to artifacts of cross-correlations with respect to auto-correlations.
At the higher probed temperatures, there was no discernible change in $S_\Sigma$ and we performed our data analysis using all the noise measurements available.

\textit{(ii) Noisy $\tau=1/2$ plateau.} A small but discernible `plateau' is present at $\tau\simeq1/2$ in the transmission versus split gate voltage of both QPC$_\mathrm{E}$ (see inset in Fig.~\ref{fig:Delta(tau)}c) and  QPC$_\mathrm{SE}$ (see Extended Data Fig.~\ref{fig:SItauSE(Vgate,nu=2/3)}).
These plateaus, which are robust to the application of a bias voltage $V$ and to temperature changes, suggest the presence of two $e^2/3h$ edge channels sequentially transmitted across the QPC. 
In that case, there would be no partition noise at the QPC, as observed at $\nu=3$ and $\nu=2/5$.
In contrast, the small $\tau\simeq1/2$ `plateaus' at $\nu=2/3$ exhibit a substantial noise signal (see also e.g.\ Ref.~\citenum{Bid_2s3noisyplateau_2009}).
It was proposed that such noise on a $\tau=1/2$ plateau was resulting not from the emergence of shot/partition noise, but from a heating mechanism involving the thermal equilibration between downstream charge modes and upstream neutral modes\cite{Mirlin_NoisyPlateaus_2020,Gefen_NoisyPlateaus_2021,Goldstein_NoisyPlateaus_2023}.
In this picture of the QPC at $\tau\sim0.5$, the fit parameters $e^*$ and $\Delta$ should not be interpreted as the charge and scaling dimension of fractional quasiparticles.
Which picture more adequately describes the QPC at $\tau\simeq1/2$ and $\nu=2/3$ is not straightforward.
On the one hand, whereas the smallness of the $\tau\simeq1/2$ `plateaus' does not rule out a simple accidental explanation within the tunneling picture (from the specific way the barrier deforms with gate voltage, possibly with nearby defects), their mere observation casts doubts on the tunneling picture and, consequently, on the interpretation of the fit parameters $e^*$ and $\Delta$ near $\tau\simeq0.5$ as characterizing quantum numbers of fractional quasiparticles.
On the other hand, the observation of similar values than for small transmissions, where the noise signal originates from the tunneling of fractional quantum Hall quasiparticles across the QPC, suggests that the same tunneling mechanism is at work at $\tau\simeq1/2$. 
In particular, a markedly larger (over-poissonian) noise would be expected from the heating interpretation in the so-called thermally equilibrated regimes (compared to $e^*/e\approx0.3$ observed here over a broad transmission range, including $\tau\simeq1/2$, and theoretically expected for the fractional quantum Hall quasiparticles at $\nu=2/3$)\cite{Mirlin_NoisyPlateaus_2020,Goldstein_NoisyPlateaus_2023}. 
Overall, more caution is advised on the interpretation of the extracted $e^*$ and $\Delta$ at $\tau\sim0.5$ for $\nu=2/3$, compared to lower $\tau$ or different $\nu\in\{1/3,2/5\}$.\\

\begin{figure*}[htb]
\renewcommand{\figurename}{\textbf{Extended Data Figure}}
	\centering
	\includegraphics[width=6cm]{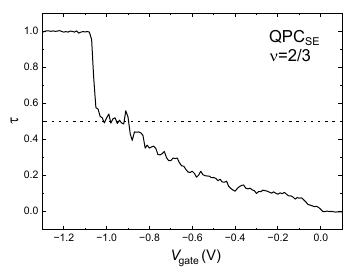}
	\caption{\textbf{Transmission `plateau' across QPC$_\mathrm{SE}$ at $\nu=2/3$.} 
The measured QPC `backscattering' transmission $\tau$ across QPC$_\mathrm{SE}$ at $\nu=2/3$ is plotted as a continuous line vs gate voltage $V_\mathrm{gate}$.
The horizontal dashed line indicates $\tau=1/2$.
See inset in Fig.~\ref{fig:Delta(tau)}c for the corresponding gate voltage sweep of QPC$_\mathrm{E}$ at $\nu=2/3$.
} 
	\label{fig:SItauSE(Vgate,nu=2/3)}
\end{figure*}

{\noindent\textbf{Extended Data.}} 
Individual values of $\Delta$ and $e^*/e$ extracted along gate voltage spans are displayed in Extended Data Figures~\ref{fig:SIDelta(tau)_1s31s6},\ref{fig:SIDelta(tau)_2s33}, in complement to Fig.~\ref{fig:Delta(tau)} of the main text.

The dc voltage dependence of the transmission $\tau(V)$ at all gate voltage tunings in the three configurations shown in Fig.~\ref{fig:Delta(tau)} are plotted in Extended Data Figure~\ref{fig:SItau(V)}. 
Among these are three $\tau(V)$ also shown in the insets of Fig.~\ref{fig:TtoSNcrossover}a,b,c.\\

\begin{figure*}[h]
\renewcommand{\figurename}{\textbf{Extended Data Figure}}
	\centering
	\includegraphics[width=18cm]{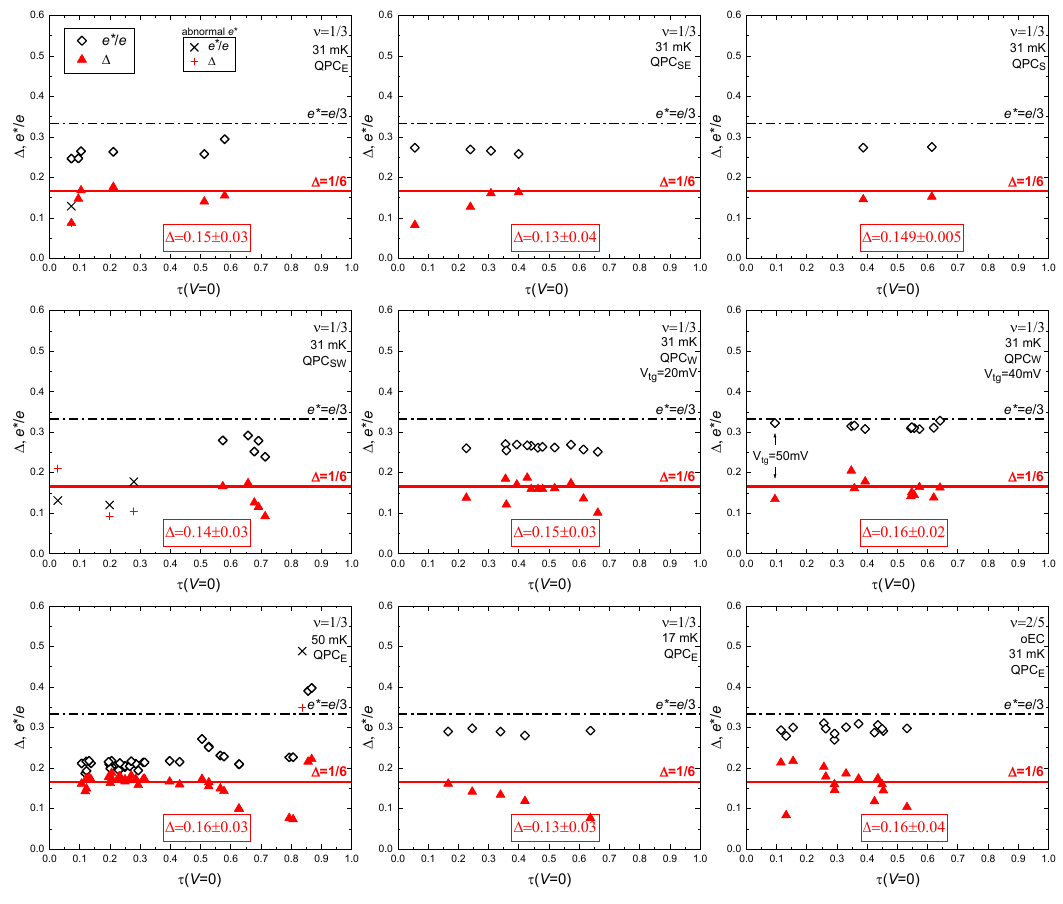}
	\caption{\textbf{Scaling dimension vs QPC tuning of predicted $\{e/3$, $\Delta=1/6\}$ quasiparticles.} 
Individual values of extracted scaling dimension ($\blacktriangle$) and charge ($\diamond$) are plotted vs $\tau(V=0)$ for each configuration addressing the predicted  $\{e/3$, $\Delta=1/6\}$ fractional quantum Hall quasiparticles.
A few points associated with anomalously low or high charge are shown as different symbols ($+$,$\times$).
Each panel corresponds to the configuration indicated within it.
The average and spread of $\Delta$ indicated in the panels are calculated only on points displayed as $\blacktriangle$ and correspond to the individual symbols with error bars in Fig.~\ref{fig:summary}.
All measurements are here at $\nu=1/3$ except the bottom right panel addressing the outer edge channel at $\nu=2/5$.
The configuration corresponding to \{$\nu=1/3$, QPC$_\mathrm{E}$, $31$\,mK, $\delta B\simeq-0.5$\,T\} is displayed in Fig.~\ref{fig:Delta(tau)}a.
} 
	\label{fig:SIDelta(tau)_1s31s6}
\end{figure*}

\begin{figure*}[h]
\renewcommand{\figurename}{\textbf{Extended Data Figure}}
	\centering
	\includegraphics[width=18cm]{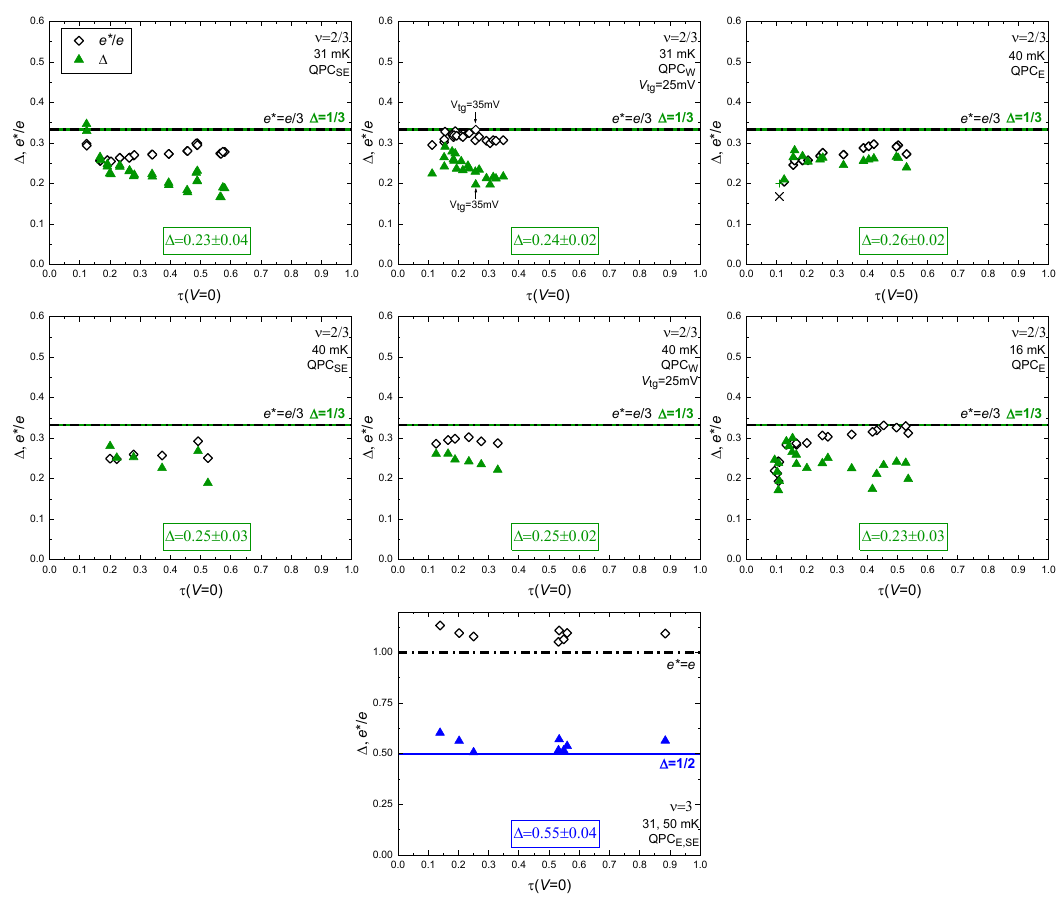}
	\caption{\textbf{Scaling dimension vs QPC tuning at $\nu=2/3$ and $\nu=3$.} 
Individual values of extracted scaling dimension $\Delta$ ($\blacktriangle$) and charge $e^*/e$ ($\diamond$) are plotted vs $\tau(V=0)$.
A few points associated with anomalously low or high charge are shown as different symbols ($+$,$\times$).
Each panel corresponds to the configuration indicated within it.
The average and spread of $\Delta$ indicated in the panels are calculated only on points displayed as $\blacktriangle$ and correspond to the individual symbols with error bars in Fig.~\ref{fig:summary}.
The six top panels correspond to $\nu=2/3$ whereas the bottom one corresponds to $\nu=3$.
The configuration corresponding to  \{$\nu=2/3$, QPC$_\mathrm{E}$, $31$\,mK\} is displayed in Fig.~\ref{fig:Delta(tau)}c.
} 
	\label{fig:SIDelta(tau)_2s33}
\end{figure*}

\begin{figure*}[ht]
\renewcommand{\figurename}{\textbf{Extended Data Figure}}
	\centering
	\includegraphics[width=18cm]{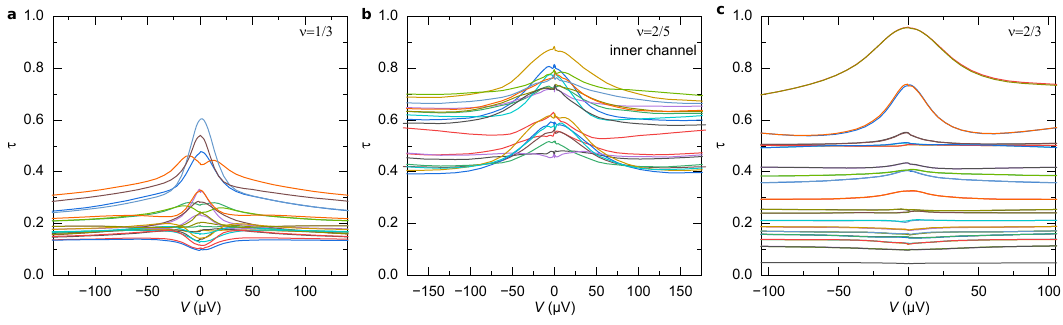}
	\caption{\textbf{Transmission vs dc voltage bias at different gate voltages.} 
The measured QPC `backscattering' transmission $\tau$ is plotted vs $V$ for the different gate voltages tunings and three QPC$_\mathrm{E}$ configurations displayed in Fig.~\ref{fig:Delta(tau)}.
Each individual tuning in each panel is shown as a line of a different color.
Panels \textbf{a,b,c} correspond to the $\nu=1/3,$ $2/5$ inner channel and $2/3$ configurations displayed in Fig.~\ref{fig:Delta(tau)}a,b,c of the main article, respectively.
} 
	\label{fig:SItau(V)}
\end{figure*}

\begin{figure*}[!htb]
\renewcommand{\figurename}{\textbf{Extended Data Figure}}
	\centering
	\includegraphics[width=6cm]{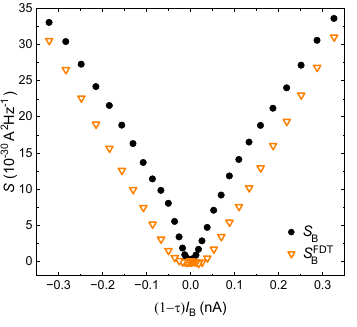}
	\caption{
	\footnotesize
 \textbf{Measured $S_\mathrm{B}$ vs calculated $S_\mathrm{B}^\mathrm{FDT}$.}
	Illustrative comparison at $\nu=1/3$ between the measured excess noise $S_\mathrm{B}$ and the value $S_\mathrm{B}^\mathrm{FDT}$ calculated from Eq.~\ref{eq:S_B^FDT} (derived with the non-equilibrium fluctuation-dissipation relation Eq.~\ref{eq:SgndT_FDT}) using the simultaneously measured $\braket{I_\mathrm{B}}(V)$.
 The noise displayed here is the same as in Fig.~\ref{fig:TtoSNcrossover}a.
 The mismatch could be explained by the same mechanism invoked for the $I-V$ characteristics (see Methods).
	}	\label{fig:TunnelingNoise}
\end{figure*}

{\noindent\textbf{Measured vs tunneling noise.}}
The measured back-scattered current $I_\mathrm{B}$ can always be written as the sum $I_\mathrm{B}=I_\mathrm{T}+I_\mathrm{gnd}$ of the incident current $I_\mathrm{gnd}$ emitted from the ohmic contact that would solely contribute to $I_\mathrm{B}$ in the absence of tunneling, and of the tunneling current $I_\mathrm{T}$ across the constriction.
With this decomposition, the back-scattered current noise reads:
\begin{equation}
    \braket{\delta I_\mathrm{B}^2}=\braket{\delta I_\mathrm{T}^2}+2\braket{\delta I_\mathrm{T}\delta I_\mathrm{gnd}}+\braket{\delta I_\mathrm{gnd}^2},
    \label{eq:SBvsST}
\end{equation}
with $\braket{\delta I_\mathrm{gnd}^2}=2k_\mathrm{B}T\nu e^2/h$ the thermal noise emitted from the grounded reservoir.
Note that since $\braket{\delta I_\mathrm{gnd}^2}$ is independent of the applied bias voltage $V$, it cancels in the excess noise $S_\mathrm{B}$. 
In the tunneling limit ($\tau_\mathrm{B}\ll1$), theory predicts from detailed balanced between upstream and downstream tunneling events that the first term in the right side of Eq.~\ref{eq:SBvsST} is independent of the scaling dimension $\Delta$ and given by \cite{Levitov_STcothDetailedBalance_2004}:
\begin{equation}
    \braket{\delta I_\mathrm{T}^2}=2e^*\braket{I_\mathrm{B}}\coth{\frac{e^*V}{2k_\mathrm{B}T}}.
    \label{eq:STT}
\end{equation}
The dependence on $\Delta$ of the measured noise thus solely results from the second term on the right side of Eq.~\ref{eq:SBvsST}, namely $2\braket{\delta I_\mathrm{T}\delta I_\mathrm{gnd}}$.
According to the so-called non-equilibrium fluctuation-dissipation relations for chiral systems (assuming a $V$-independent Hamiltonian, as discussed below), this $\Delta$ dependent contribution to the noise is simply given by \cite{Wang_nonEqFDTprb_2011}:
\begin{equation}
    \braket{\delta I_\mathrm{T}\delta I_\mathrm{gnd}}=-2k_\mathrm{B}T\frac{\partial \braket{I_\mathrm{B}}}{\partial V}.
    \label{eq:SgndT_FDT}
\end{equation}
Experimentally, $\partial\braket{I_\mathrm{B}}/\partial V$ is directly measured.
Hence, in this framework, one could calculate the excess noise $S_\mathrm{B}^\mathrm{FDT}$ by plugging the separately measured tunneling current and its derivative into these equations.
This gives (with in addition the usual ad hoc correction for large $\tau$)
\begin{equation}
    S_\mathrm{B}^\mathrm{FDT}=2e^*(1-\tau)\braket{I_\mathrm{B}}\coth{\frac{e^*V}{2k_\mathrm{B}T}}-4k_\mathrm{B}T\left(1-\frac{\partial \braket{I_\mathrm{B}}}{\partial I_\mathrm{inj}}\right)\frac{\partial \braket{I_\mathrm{B}}}{\partial V}.
    \label{eq:S_B^FDT}
\end{equation}
However, as illustrated in Extended Data Fig.~\ref{fig:TunnelingNoise}, we find that Eq.~\ref{eq:S_B^FDT} using the measured $\braket{I_\mathrm{B}}(V)$ does not reproduce the simultaneously measured thermal to shot noise crossover.
This should not come as a surprise since the current and its derivative do not follow Luttinger liquid predictions.
One could explain this mismatch by invoking the same possible explanation as for the data-theory discrepancy on the $I-V$ characteristics, namely that the shape of the QPC potential, and therefore the quasiparticle tunneling amplitude, is impacted by external parameters, such as an electrostatic deformation induced by a change in the applied bias voltage, the temperature or the tunneling current itself \cite{Shtanko_SN2s3ExtParam_2014}.  
Indeed, as pointed-out in \cite{Wang_nonEqFDTprb_2011}, Eq.~\ref{eq:SgndT_FDT} holds if the voltage bias $V$ only manifests through the chemical potential of the incident edge channel, and not if applying $V$ impacts the tunnel Hamiltonian.

\begin{thebibliography}{62}
\expandafter\ifx\csname natexlab\endcsname\relax\def\natexlab#1{#1}\fi
\expandafter\ifx\csname url\endcsname\relax
  \def\url#1{\texttt{#1}}\fi
\expandafter\ifx\csname urlprefix\endcsname\relax\def\urlprefix{URL }\fi

\bibitem[{Wen(2004)}]{Wen_BookQFT_2004}
Wen, X.~G.
\newblock \emph{Quantum Field Theory of Many-Body Systems: From the Origin of
  Sound to an Origin of Light and Electrons} (Oxford University Press, Oxford,
  2004).

\bibitem[{Jain(2007)}]{Jain_BookCompositeFermions_2007}
Jain, J.
\newblock \emph{Composite Fermions} (Cambridge University Press, Cambridge,
  2007).

\bibitem[{Goldman \& Su(1995)}]{Goldman_FracQAntiDot_1995}
Goldman, V.~J. \& Su, B.
\newblock Resonant Tunneling in the Quantum Hall Regime: Measurement of
  Fractional Charge.
\newblock \emph{Science} \textbf{267}, 1010--1012 (1995).

\bibitem[{de~Picciotto \emph{et~al.}(1997)}]{de-Picciotto_es3_1997}
de~Picciotto, R. \emph{et~al.}
\newblock Direct observation of a fractional charge.
\newblock \emph{Nature} \textbf{389}, 162--164 (1997).

\bibitem[{Saminadayar \emph{et~al.}(1997)Saminadayar, Glattli, Jin \&
  Etienne}]{Saminadayar_FracE1s3_1997}
Saminadayar, L., Glattli, D.~C., Jin, Y. \& Etienne, B.
\newblock Observation of the $\mathit{e}\mathit{/}3$ Fractionally Charged
  Laughlin Quasiparticle.
\newblock \emph{Phys. Rev. Lett.} \textbf{79}, 2526--2529 (1997).

\bibitem[{Nakamura \emph{et~al.}(2020)Nakamura, Liang, Gardner \&
  Manfra}]{Nakamura_Anyon1s3_2020}
Nakamura, J., Liang, S., Gardner, G.~C. \& Manfra, M.~J.
\newblock Direct observation of anyonic braiding statistics.
\newblock \emph{Nat. Phys.} \textbf{16}, 931--936 (2020).

\bibitem[{Bartolomei \emph{et~al.}(2020)}]{Bartolomei_Cross_1tiers_2020}
Bartolomei, H. \emph{et~al.}
\newblock Fractional statistics in anyon collisions.
\newblock \emph{Science} \textbf{368}, 173--177 (2020).

\bibitem[{Rosenow \& Halperin(2002)}]{Rosenow_NonUnivQPCfqhe_2004}
Rosenow, B. \& Halperin, B.~I.
\newblock Nonuniversal Behavior of Scattering between Fractional Quantum Hall
  Edges.
\newblock \emph{Phys. Rev. Lett.} \textbf{88}, 096404 (2002).

\bibitem[{Papa \& MacDonald(2004)}]{Papa_anomTDOSinteraction_2004}
Papa, E. \& MacDonald, A.~H.
\newblock Interactions Suppress Quasiparticle Tunneling at Hall Bar
  Constrictions.
\newblock \emph{Phys. Rev. Lett.} \textbf{93}, 126801 (2004).

\bibitem[{Shtanko \emph{et~al.}(2014)Shtanko, Snizhko \&
  Cheianov}]{Shtanko_SN2s3ExtParam_2014}
Shtanko, O., Snizhko, K. \& Cheianov, V.
\newblock Nonequilibrium noise in transport across a tunneling contact between
  $\ensuremath{\nu}=\frac{2}{3}$ fractional quantum Hall edges.
\newblock \emph{Phys. Rev. B} \textbf{89}, 125104 (2014).

\bibitem[{Dolcetto \emph{et~al.}(2012)Dolcetto, Barbarino, Ferraro, Magnoli \&
  Sassetti}]{Dolcetto_TDOSextendedQPC_2012}
Dolcetto, G., Barbarino, S., Ferraro, D., Magnoli, N. \& Sassetti, M.
\newblock Tunneling between helical edge states through extended contacts.
\newblock \emph{Phys. Rev. B} \textbf{85}, 195138 (2012).

\bibitem[{Snizhko \& Cheianov(2015)}]{Snizhko_ScaldimFano_2015}
Snizhko, K. \& Cheianov, V.
\newblock Scaling dimension of quantum Hall quasiparticles from
  tunneling-current noise measurements.
\newblock \emph{Phys. Rev. B} \textbf{91}, 195151 (2015).

\bibitem[{Schiller \emph{et~al.}(2022)Schiller, Oreg \&
  Snizhko}]{Schiller_ScaldimFano_2022}
Schiller, N., Oreg, Y. \& Snizhko, K.
\newblock Extracting the scaling dimension of quantum Hall quasiparticles from
  current correlations.
\newblock \emph{Phys. Rev. B} \textbf{105}, 165150 (2022).

\bibitem[{Heiblum(2010)}]{Heiblum_FracChargeMeas_2010}
Heiblum, M.
\newblock \emph{Fractional Charge Determination via Quantum Shot Noise
  Measurements}, 115--136 (in Perspectives of Mesoscopic Physics, Edited by A.
  Aharony and O. Antin-Wholman, World Scientific Publishing, 2010).

\bibitem[{Nayak \emph{et~al.}(2008)Nayak, Simon, Stern, Freedman \&
  Das~Sarma}]{Nayak_TopoRMP_2008}
Nayak, C., Simon, S.~H., Stern, A., Freedman, M. \& Das~Sarma, S.
\newblock Non-Abelian anyons and topological quantum computation.
\newblock \emph{Rev. Mod. Phys.} \textbf{80}, 1083--1159 (2008).

\bibitem[{Giamarchi(2004)}]{Giamarchi_TLLbook_2004}
Giamarchi, T.
\newblock \emph{Quantum Physics in One Dimension}.
\newblock International Series of Monographs on Physics (Oxford University
  Press, 2004).

\bibitem[{Jonckheere \emph{et~al.}(2023)Jonckheere, Rech, Gr\'emaud \&
  Martin}]{Jonckheere_HOMdelta_2023}
Jonckheere, T., Rech, J., Gr\'emaud, B. \& Martin, T.
\newblock Anyonic Statistics Revealed by the Hong-Ou-Mandel Dip for Fractional
  Excitations.
\newblock \emph{Phys. Rev. Lett.} \textbf{130}, 186203 (2023).

\bibitem[{Iyer \emph{et~al.}(2024)}]{Iyer_FiniteWidthAnyons_2024}
Iyer, K. \emph{et~al.}
\newblock Finite Width of Anyons Changes Their Braiding Signature.
\newblock \emph{Phys. Rev. Lett.} \textbf{132}, 216601 (2024).

\bibitem[{Thamm \& Rosenow(2024)}]{Thamm_finitewidthanyons_2024}
Thamm, M. \& Rosenow, B.
\newblock Effect of the Soliton Width on Nonequilibrium Exchange Phases of
  Anyons.
\newblock \emph{Phys. Rev. Lett.} \textbf{132}, 156501 (2024).

\bibitem[{Reznikov \emph{et~al.}(1999)Reznikov, de~Picciotto, Griffiths,
  Heiblum \& Umansky}]{Reznikov_es5_1999}
Reznikov, M., de~Picciotto, R., Griffiths, T., Heiblum, M. \& Umansky, V.
\newblock Observation of quasiparticles with one-fifth of an electron's charge.
\newblock \emph{Nature} \textbf{399}, 238--241 (1999).

\bibitem[{Martin \emph{et~al.}(2004)}]{Martin_SetFracChargeFQHE_2004}
Martin, J. \emph{et~al.}
\newblock Localization of Fractionally Charged Quasi-Particles.
\newblock \emph{Science} \textbf{305}, 980--983 (2004).

\bibitem[{Dolev \emph{et~al.}(2008)Dolev, Heiblum, Umansky, Stern \&
  Mahalu}]{Dolev_nu5s2SNes4_2008}
Dolev, M., Heiblum, M., Umansky, V., Stern, A. \& Mahalu, D.
\newblock Observation of a quarter of an electron charge at the $\nu$ = 5/2
  quantum Hall state.
\newblock \emph{Nature} \textbf{452}, 829--834 (2008).

\bibitem[{Venkatachalam \emph{et~al.}(2011)Venkatachalam, Yacoby, Pfeiffer \&
  West}]{Venkatachalam_SetChargees4Nu5s2_2011}
Venkatachalam, V., Yacoby, A., Pfeiffer, L. \& West, K.
\newblock Local charge of the $\nu$ = 5/2 fractional quantum Hall state.
\newblock \emph{Nature} \textbf{469}, 185–188 (2011).

\bibitem[{Kapfer \emph{et~al.}(2019)}]{Kapfer_FQHE_2019}
Kapfer, M. \emph{et~al.}
\newblock A Josephson relation for fractionally charged anyons.
\newblock \emph{Science} \textbf{363}, 846--849 (2019).

\bibitem[{Bisognin \emph{et~al.}(2019)}]{Bisognin_PATes3_2019}
Bisognin, R. \emph{et~al.}
\newblock Microwave photons emitted by fractionally charged quasiparticles.
\newblock \emph{Nature Communications} \textbf{10}, 1708 (2019).

\bibitem[{Röösli \emph{et~al.}(2021)}]{Roosli_fracCB_2021}
Röösli, M.~P. \emph{et~al.}
\newblock Fractional Coulomb blockade for quasi-particle tunneling between edge
  channels.
\newblock \emph{Science Advances} \textbf{7}, eabf5547 (2021).

\bibitem[{Nakamura \emph{et~al.}(2023)Nakamura, Liang, Gardner \&
  Manfra}]{Nakamura_Anyon2s5_2023}
Nakamura, J., Liang, S., Gardner, G.~C. \& Manfra, M.~J.
\newblock Fabry-P\'erot Interferometry at the $\ensuremath{\nu}=2/5$ Fractional
  Quantum Hall State.
\newblock \emph{Phys. Rev. X} \textbf{13}, 041012 (2023).

\bibitem[{Glidic \emph{et~al.}(2023{\natexlab{a}})}]{Glidic_Collider2023}
Glidic, P. \emph{et~al.}
\newblock Cross-Correlation Investigation of Anyon Statistics in the
  $\ensuremath{\nu}=1/3$ and $2/5$ Fractional Quantum Hall States.
\newblock \emph{Phys. Rev. X} \textbf{13}, 011030 (2023{\natexlab{a}}).

\bibitem[{Ruelle \emph{et~al.}(2023)}]{Ruelle_2s5_2023}
Ruelle, M. \emph{et~al.}
\newblock Comparing Fractional Quantum Hall Laughlin and Jain Topological
  Orders with the Anyon Collider.
\newblock \emph{Phys. Rev. X} \textbf{13}, 011031 (2023).

\bibitem[{Lee \emph{et~al.}(2023)}]{Lee_BunchingOrBraiding_2023}
Lee, J.-Y.~M. \emph{et~al.}
\newblock Partitioning of diluted anyons reveals their braiding statistics.
\newblock \emph{Nature} \textbf{617}, 281 (2023).

\bibitem[{Rosenow \emph{et~al.}(2016)Rosenow, Levkivskyi \&
  Halperin}]{Rosenow_Collider_2016}
Rosenow, B., Levkivskyi, I.~P. \& Halperin, B.~I.
\newblock Current Correlations from a Mesoscopic Anyon Collider.
\newblock \emph{Phys. Rev. Lett.} \textbf{116}, 156802 (2016).

\bibitem[{Lee \& Sim(2022)}]{Lee_NonAbelianCollider_2022}
Lee, J.-Y.~M. \& Sim, H.-S.
\newblock Non-Abelian anyon collider.
\newblock \emph{Nature Communications} \textbf{13}, 6660 (2022).

\bibitem[{Roddaro \emph{et~al.}(2004)Roddaro, Pellegrini, Beltram, Biasiol \&
  Sorba}]{Roddaro_WBSscaldim_2004}
Roddaro, S., Pellegrini, V., Beltram, F., Biasiol, G. \& Sorba, L.
\newblock Interedge Strong-to-Weak Scattering Evolution at a Constriction in
  the Fractional Quantum Hall Regime.
\newblock \emph{Phys. Rev. Lett.} \textbf{93}, 046801 (2004).

\bibitem[{Radu \emph{et~al.}(2008)}]{Radu_TDOS5s2_2008}
Radu, I.~P. \emph{et~al.}
\newblock Quasi-particle properties from Tunneling in the $\nu$ = 5/2
  Fractional Quantum Hall State.
\newblock \emph{Science} \textbf{320}, 899 (2008).

\bibitem[{Baer \emph{et~al.}(2014)}]{Baer_TDOS5s2_2014}
Baer, S. \emph{et~al.}
\newblock Experimental probe of topological orders and edge excitations in the
  second Landau level.
\newblock \emph{Phys. Rev. B} \textbf{90}, 075403 (2014).

\bibitem[{Anthore \emph{et~al.}(2018)}]{Anthore_QuSimTLL_2018}
Anthore, A. \emph{et~al.}
\newblock Circuit Quantum Simulation of a Tomonaga-Luttinger Liquid with an
  Impurity.
\newblock \emph{Phys. Rev. X} \textbf{8}, 031075 (2018).

\bibitem[{Cohen \emph{et~al.}(2023)}]{Cohen_UnivCLL1s3to1_2023}
Cohen, L.~A. \emph{et~al.}
\newblock Universal chiral Luttinger liquid behavior in a graphene fractional
  quantum Hall point contact.
\newblock \emph{Science} \textbf{382}, 542--547 (2023).

\bibitem[{Chang(2003)}]{Chang_LL_2003}
Chang, A.~M.
\newblock Chiral Luttinger liquids at the fractional quantum Hall edge.
\newblock \emph{Rev. Mod. Phys.} \textbf{75}, 1449--1505 (2003).

\bibitem[{Lin \emph{et~al.}(2012)Lin, Dillard, Kastner, Pfeiffer \&
  West}]{Lin_TDOS5s2_2012}
Lin, X., Dillard, C., Kastner, M.~A., Pfeiffer, L.~N. \& West, K.~W.
\newblock Measurements of quasiparticle tunneling in the
  $\ensuremath{\upsilon}=\frac{5}{2}$ fractional quantum Hall state.
\newblock \emph{Phys. Rev. B} \textbf{85}, 165321 (2012).

\bibitem[{Yang(2003)}]{Yang_TDOSwithEdgeReconstruct_2003}
Yang, K.
\newblock Field Theoretical Description of Quantum Hall Edge Reconstruction.
\newblock \emph{Phys. Rev. Lett.} \textbf{91}, 036802 (2003).

\bibitem[{Ferraro \emph{et~al.}(2008)Ferraro, Braggio, Merlo, Magnoli \&
  Sassetti}]{Ferraro_AnomTunnelNeutralModes_2008}
Ferraro, D., Braggio, A., Merlo, M., Magnoli, N. \& Sassetti, M.
\newblock Relevance of Multiple Quasiparticle Tunneling between Edge States at
  $\ensuremath{\nu}=p/(2np+1)$.
\newblock \emph{Phys. Rev. Lett.} \textbf{101}, 166805 (2008).

\bibitem[{Rech \emph{et~al.}(2020)Rech, Jonckheere, Gr\'emaud \&
  Martin}]{Rech_deltaTnoiseFQHE_2020}
Rech, J., Jonckheere, T., Gr\'emaud, B. \& Martin, T.
\newblock Negative Delta-$T$ Noise in the Fractional Quantum Hall Effect.
\newblock \emph{Phys. Rev. Lett.} \textbf{125}, 086801 (2020).

\bibitem[{Zhang \emph{et~al.}(2022)Zhang, Gornyi \&
  Sp\aa{}nsl\"att}]{Zhang_deltaTnoiseFQHE_2022}
Zhang, G., Gornyi, I.~V. \& Sp\aa{}nsl\"att, C.
\newblock Delta-$T$ noise for weak tunneling in one-dimensional systems:
  Interactions versus quantum statistics.
\newblock \emph{Phys. Rev. B} \textbf{105}, 195423 (2022).

\bibitem[{Ebisu \emph{et~al.}(2022)Ebisu, Schiller \&
  Oreg}]{Ebisu_deltaGQvsDelta_2022}
Ebisu, H., Schiller, N. \& Oreg, Y.
\newblock Fluctuations in Heat Current and Scaling Dimension.
\newblock \emph{Phys. Rev. Lett.} \textbf{128}, 215901 (2022).

\bibitem[{Kane \& Fisher(1995)}]{Kane_ImpScattFQHE_1995}
Kane, C.~L. \& Fisher, M. P.~A.
\newblock Impurity scattering and transport of fractional quantum Hall edge
  states.
\newblock \emph{Phys. Rev. B} \textbf{51}, 13449--13466 (1995).

\bibitem[{Bid \emph{et~al.}(2009)Bid, Ofek, Heiblum, Umansky \&
  Mahalu}]{Bid_2s3noisyplateau_2009}
Bid, A., Ofek, N., Heiblum, M., Umansky, V. \& Mahalu, D.
\newblock Shot Noise and Charge at the $2/3$ Composite Fractional Quantum Hall
  State.
\newblock \emph{Phys. Rev. Lett.} \textbf{103}, 236802 (2009).

\bibitem[{Blanter \& Büttiker(2000)}]{blanter_shotnoise_2000}
Blanter, Y. \& Büttiker, M.
\newblock Shot noise in mesoscopic conductors.
\newblock \emph{Phys. Rep.} \textbf{336}, 1--166 (2000).

\bibitem[{Griffiths \emph{et~al.}(2000)Griffiths, Comforti, Heiblum, Stern \&
  Umansky}]{Griffiths_EstarVsTau_2000}
Griffiths, T.~G., Comforti, E., Heiblum, M., Stern, A. \& Umansky, V.
\newblock Evolution of Quasiparticle Charge in the Fractional Quantum Hall
  Regime.
\newblock \emph{Phys. Rev. Lett.} \textbf{85}, 3918--3921 (2000).

\bibitem[{Davies \& Larkin(1994)}]{Davis_GateGaAsStrain_1994}
Davies, J.~H. \& Larkin, I.~A.
\newblock Theory of potential modulation in lateral surface superlattices.
\newblock \emph{Phys. Rev. B} \textbf{49}, 4800--4809 (1994).

\bibitem[{Glidic \emph{et~al.}(2023{\natexlab{b}})}]{Glidic_Andreev_2023}
Glidic, P. \emph{et~al.}
\newblock Quasiparticle Andreev scattering in the $\ensuremath{\nu}=1/3$
  fractional quantum Hall regime.
\newblock \emph{Nature Communications} \textbf{14}, 514 (2023{\natexlab{b}}).

\bibitem[{Kamata \emph{et~al.}(2014)Kamata, Kumada, Hashisaka, Muraki \&
  Fujisawa}]{Kamata_artificialTLL_2014}
Kamata, H., Kumada, N., Hashisaka, M., Muraki, K. \& Fujisawa, T.
\newblock Fractionalized wave packets from an artificial Tomonaga–Luttinger
  liquid.
\newblock \emph{Nature Nanotechnology} \textbf{9}, 177–181 (2014).

\bibitem[{Iftikhar \emph{et~al.}(2016)}]{Iftikhar_Ttriad_2016}
Iftikhar, Z. \emph{et~al.}
\newblock Primary thermometry triad at 6 mK in mesoscopic circuits.
\newblock \emph{Nat. Commun.} \textbf{7}, 12908 (2016).

\bibitem[{Liang \emph{et~al.}(2012)Liang, Dong, Gennser, Cavanna \&
  Jin}]{Liang_HEMTs_2012}
Liang, Y., Dong, Q., Gennser, U., Cavanna, A. \& Jin, Y.
\newblock Input Noise Voltage Below 1\,nV/Hz$^{1/2}$ at 1\,kHz in the HEMTs at
  4.2\,K.
\newblock \emph{J. Low Temp. Phys.} \textbf{167}, 632--637 (2012).

\bibitem[{Jezouin \emph{et~al.}(2013)}]{Jezouin_QLimHeatFlow_2013b}
Jezouin, S. \emph{et~al.}
\newblock Quantum Limit of Heat Flow Across a Single Electronic Channel.
\newblock \emph{Science} \textbf{342}, 601--604 (2013).

\bibitem[{Batra \& Feldman(2024)}]{Barta_AutovsCross_2023}
Batra, N. \& Feldman, D.~E.
\newblock Different Fractional Charges from Auto- and Cross-Correlation Noise
  in Quantum Hall States without Upstream Modes.
\newblock \emph{Phys. Rev. Lett.} \textbf{132}, 226601 (2024).

\bibitem[{Kane \emph{et~al.}(1994)Kane, Fisher \&
  Polchinski}]{Kane_2s3disorder_1994}
Kane, C.~L., Fisher, M. P.~A. \& Polchinski, J.
\newblock Randomness at the edge: Theory of quantum Hall transport at filling
  \ensuremath{\nu}=2/3.
\newblock \emph{Phys. Rev. Lett.} \textbf{72}, 4129--4132 (1994).

\bibitem[{Gross \emph{et~al.}(2012)Gross, Dolev, Heiblum, Umansky \&
  Mahalu}]{Gross_UpstreamNeutral2s3}
Gross, Y., Dolev, M., Heiblum, M., Umansky, V. \& Mahalu, D.
\newblock Upstream Neutral Modes in the Fractional Quantum Hall Effect Regime:
  Heat Waves or Coherent Dipoles.
\newblock \emph{Phys. Rev. Lett.} \textbf{108}, 226801 (2012).

\bibitem[{Sp\aa{}nsl\"att \emph{et~al.}(2020)Sp\aa{}nsl\"att, Park, Gefen \&
  Mirlin}]{Mirlin_NoisyPlateaus_2020}
Sp\aa{}nsl\"att, C., Park, J., Gefen, Y. \& Mirlin, A.~D.
\newblock Conductance plateaus and shot noise in fractional quantum Hall point
  contacts.
\newblock \emph{Phys. Rev. B} \textbf{101}, 075308 (2020).

\bibitem[{Park \emph{et~al.}(2021)Park, Rosenow \&
  Gefen}]{Gefen_NoisyPlateaus_2021}
Park, J., Rosenow, B. \& Gefen, Y.
\newblock Symmetry-related transport on a fractional quantum Hall edge.
\newblock \emph{Phys. Rev. Res.} \textbf{3}, 023083 (2021).

\bibitem[{Manna \emph{et~al.}(2024)Manna, Das, Gefen \&
  Goldstein}]{Goldstein_NoisyPlateaus_2023}
Manna, S., Das, A., Gefen, Y. \& Goldstein, M.
\newblock Shot noise as a diagnostic in the $\nu=2/3$ fractional quantum Hall
  edge zoo (2024).
\newblock ArXiv: 2307.05175.

\bibitem[{Levitov \& Reznikov(2004)}]{Levitov_STcothDetailedBalance_2004}
Levitov, L.~S. \& Reznikov, M.
\newblock Counting statistics of tunneling current.
\newblock \emph{Phys. Rev. B} \textbf{70}, 115305 (2004).

\bibitem[{Wang \& Feldman(2011)}]{Wang_nonEqFDTprb_2011}
Wang, C. \& Feldman, D.~E.
\newblock Fluctuation-dissipation theorem for chiral systems in nonequilibrium
  steady states.
\newblock \emph{Phys. Rev. B} \textbf{84}, 235315 (2011).

\end{thebibliography}
\end{document}